\documentclass[12pt]{article}

\usepackage{amsmath,amssymb,amsfonts,amscd,mathrsfs,mathtools,bbold, braket}
\usepackage{xcolor}
\definecolor{darkblue}{rgb}{0.1,0.1,.7}
\usepackage[colorlinks, linkcolor=darkblue, citecolor=darkblue, urlcolor=darkblue, linktocpage]{hyperref} 
\usepackage{geometry}
\usepackage{subcaption}
\usepackage{tablefootnote,colortbl}

\usepackage{ physics }

\usepackage{braket}
\usepackage{cancel}
\usepackage{arcs}
\usepackage{bm}
\usepackage{arydshln}
\usepackage{booktabs,multirow}
\usepackage{amssymb}
\usepackage{fancyvrb}

\usepackage{tikz-feynman}
 \tikzfeynmanset{compat=1.0.0}

 \usepackage{mcite}
 
 \usepackage[square, comma, compress,numbers]{natbib}
\usepackage{subcaption}
\usepackage[utf8]{inputenc} 

\definecolor{myorange}{RGB}{199,146,32}

\definecolor{Gray1}{gray}{0.97}
\definecolor{Gray2}{gray}{0.9}
\definecolor{LightCyan}{rgb}{0.88,1,1}
\definecolor{blu}{rgb}{0,0,1}

\usepackage{ragged2e}
\usepackage{array}
\newcolumntype{L}[1]{>{\raggedright\let\newline\\\arraybackslash\hspace{0pt}}m{#1}}
\newcolumntype{C}[1]{>{\centering\let\newline\\\arraybackslash\hspace{0pt}}m{#1}}
\newcolumntype{R}[1]{>{\raggedleft\let\newline\\\arraybackslash\hspace{0pt}}m{#1}}

\usepackage{dsfont} 
\usepackage{tcolorbox}
\usepackage{booktabs}

\usepackage{titlesec}
\titleformat*{\section}{\large\bfseries}
\titleformat*{\subsection}{\normalsize\bfseries}
\titleformat*{\subsubsection}{\normalsize\it}
\titleformat*{\paragraph}{\normalsize\bfseries}
\titleformat*{\subparagraph}{\normalsize\bfseries}

\newcommand{\reef}[1]{(\ref{#1})}

\newcommand{\la}{\langle}
\newcommand{\ra}{\rangle}

\newcommand{\beq}{\begin{equation}} 
\newcommand{\eeq}{\end{equation}}

\def\geq{\geqslant}

\newcommand{\diffop}[2]{\ifthenelse{\equal{#2}{1}}{\frac{\mrm{d}}{\mrm{d} #1}}{\frac{\mrm{d}^#2}{\mrm{d} #1^#2}}}

\newcommand{\mrm}[1]{{\mathrm #1}}

\usepackage[normalem]{ulem}

\DeclareMathOperator*{\SumInt}{%
\mathchoice%
  {\ooalign{$\displaystyle\sum$\cr\hidewidth$\displaystyle\int$\hidewidth\cr}}
  {\ooalign{\raisebox{.14\height}{\scalebox{.7}{$\textstyle\sum$}}\cr\hidewidth$\textstyle\int$\hidewidth\cr}}
  {\ooalign{\raisebox{.2\height}{\scalebox{.6}{$\scriptstyle\sum$}}\cr$\scriptstyle\int$\cr}}
  {\ooalign{\raisebox{.2\height}{\scalebox{.6}{$\scriptstyle\sum$}}\cr$\scriptstyle\int$\cr}}
}

\newcommand{\be}{\begin{equation}}
\newcommand{\ee}{\end{equation}}
\def\bea#1\eea{\begin{align}#1\end{align}}

\newcommand{\eq}[1]{Eq.~(\ref{#1})}

\definecolor{Ecolor}{RGB}{106,157,235}

\usepackage{accents}
\newlength{\dhatheight}

\usepackage{tikz}
\usetikzlibrary{arrows,shapes}
\usetikzlibrary{trees}
\usetikzlibrary{matrix,arrows} 				
\usetikzlibrary{calc} 
\usetikzlibrary{positioning}				
\usetikzlibrary{calc,through}				
\usetikzlibrary{decorations.pathreplacing}  
\usepackage[tikz]{bclogo} 					
\usepackage{pgffor}							

\usetikzlibrary{decorations.pathmorphing}	
\usetikzlibrary{decorations.markings}
\usetikzlibrary{intersections}				
\tikzset{
    vector/.style={decorate, decoration={snake}, draw},
    graviton/.style={decorate, decoration={snake,amplitude=1.5pt}, draw},
    fermion/.style={postaction={decorate},
        decoration={markings,mark=at position .55 with {\arrow{>}}}},
    fermionbar/.style={draw, postaction={decorate},
        decoration={markings,mark=at position .55 with {\arrow{<}}}},
    fermionnoarrow/.style={},
    gluon/.style={decorate,
        decoration={coil,amplitude=4pt, segment length=5pt}},
    scalar/.style={dashed, postaction={decorate},
        decoration={markings,mark=at position .55 with {\arrow{>}}}},
    scalarbar/.style={dashed, postaction={decorate},
        decoration={markings,mark=at position .55 with {\arrow{<}}}},
    scalarnoarrow/.style={dashed,draw},
	provector/.style={decorate, decoration={snake,amplitude=2.5pt}, draw},
	antivector/.style={decorate, decoration={snake,amplitude=-2.5pt}, draw},
	    electron/.style={draw=black, postaction={decorate},
        decoration={markings,mark=at position .55 with {\arrow[draw=black]{>}}}},
	bigvector/.style={decorate, decoration={snake,amplitude=4pt}, draw},
	vectorscalar/.style={loosely dotted,draw=black, postaction={decorate}},
}

\numberwithin{equation}{section}
\usepackage{tocloft}
\begin{document}

\begin{center}

\vspace{2.0cm}
{\Large \bf 
{Thermodynamics from the S-matrix reloaded: \\[.2cm] emergent thermal mass }
}

\vspace{1.0cm}
{\small 
Pietro Baratella$^{(a)}$ and 
Joan Elias Mir\'o$^{(b)}$
}

\vspace{.3cm}

$(a)$ \emph{Jo\v{z}ef Stefan Institute, Jamova 39, 1000 Ljubljana, Slovenia}

$(b)$ \emph{ICTP, Str. Costiera, 11, 34151, Trieste, Italy}

\vspace{0.6cm}

\abstract{\noindent 
The formalism of Dashen, Ma and Bernstein (DMB) expresses the thermal partition function  of a system in terms of the S-matrix operator, roughly
$
Z(\beta) \propto \int dE\, e^{-\beta E}\,\mathrm{Tr}\,\ln S(E),
$
where  $S$ denotes the full scattering operator on the asymptotic Fock space -- i.e. including all multi-particle sectors -- defined via the Lippmann–Schwinger equation.
Recently we have employed this formalism to compute the free energy of flux tubes (essentially a two-dimensional theory of derivatively coupled scalars) and the two-loop $O(\alpha_s)$ QCD thermal free energy.

Moving to higher orders, it is well known that at $O(\alpha_s^2)$ in QCD, or e.g. at  $O(\lambda^2)$ in $\lambda\phi^4$ theory,  the free energy develops IR divergences. These IR divergences are resolved  by the screening Debye mass.
However, the DMB formalism expresses the free energy in terms of a trace of the S-matrix operator in the vacuum. 
How, then,  does the Debye mass arise in this framework? In this work we address this question, thereby paving the way for higher-order applications of the DMB formalism in relativistic QFT.

} 
\end{center}
\setcounter{footnote}{0}

\vspace{.5cm}

\setcounter{tocdepth}{2}
\tableofcontents

\section{Introduction}

Our current understanding of elementary particle interactions is encoded in the Standard Model Lagrangian, from which zero-temperature physics — such as vacuum scattering amplitudes — is extracted using the methods of Quantum Field Theory (QFT).
The same elementary degrees of freedom that make up the Standard Model were once part of a cosmic plasma at a finite temperature. 
To study this and for many other applications, ranging from condensed matter to heavy-ion collisions and astrophysics, one needs the tools of thermal QFT (Th-QFT).

The standard approach to perturbative  Th-QFT instructs us to compute Feynman  diagrams in Euclidean signature with compact time and periodic boundary conditions.
A less-traveled route to approach Th-QFT is the formalism of Dashen, Ma, and Bernstein (DMB) \cite{Dashen:1969ep}, which expresses the thermal free energy of a system succinctly in terms of the S-matrix operator:
\be\label{master}
F=F_0-\frac 1{2\pi i}\int_0^\infty \dd E\, e^{-\beta E}\, {\rm Tr}_c\ln S(E)\,,
\ee
where $F_0$ is the free energy of the non-interacting asymptotic states, while $S(E)$ is an operator-valued distribution in the energy variable $E$ that extends off-shell the familiar S-matrix operator, in the sense that
\be\label{defSE}
\la b |S(E\equiv E_a)|a \ra=S_{ba}\,,
\ee
where $a,b$ are asymptotic states, $E_a$ is the energy of state $a$ and $S_{ba}$ is the usual S-matrix element for the scattering $a\to b$. The off-shell S-matrix is defined as $S(E)=1+\big(G(E)-\bar G(E)\big)T(E)$, where $G(E)=(E-H_0+i\varepsilon)^{-1}$ is the the free-theory resolvent --- defined in terms of the free Hamiltonian $H_0$, $\bar G(E)$ its complex conjugate and
\be\label{Lipp}
T(E)=V+V\,\frac1{E-H_0+i \varepsilon}\,V+V\,\frac1{E-H_0+i \varepsilon}\,V\,\frac1{E-H_0+i \varepsilon}\,V+\ldots
\ee
with $V=H-H_0$ the interaction Hamiltonian. Taking $\lim_{E\to E_a}\la b|T(E)|a\ra$ gives the Lippmann-Schwinger prescription for computing perturbatively the amplitude for the process $a\to b$.
The logarithm `ln' of the off-shell scattering operator which appears in \reef{master} is defined via its formal Taylor series about the identity operator.~\footnote{
Eq.~\reef{master} is valid in any spacetime dimension and involves the entire S-matrix operator, which connects all $n$-to-$n$ particles sectors. In integrable systems in $(1+1)$-dimensions the Thermodyinamic Bethe Ansatz (TBA) expresses the  free-energy in terms of the two-to-two S-matrix only. See refs.~\cite{Schubring:2024yfi,Baratella:2024sax} for a comparison between DMB and TBA.
}

The past decades have seen remarkable progress in our understanding of scattering amplitudes and the development of sophisticated techniques for their evaluation. Building on this momentum, we recently initiated a systematic revision of DMB methods in ref.~\cite{Baratella:2024sax}, see also ref.~\cite{Schubring:2024yfi}.

The motivation for this program is manifold, particularly within the context of QCD.
On the theoretical side, on-shell methods have emerged as exceptionally powerful tools for describing gauge theories, often bypassing the redundancies inherent in traditional Lagrangian approaches. From the computational viewpoint, there is now a vast repository of knowledge regarding the analytic structure of QCD amplitudes and their supersymmetric counterparts. Finally, interest in the properties of the quark–gluon plasma, which describes matter at temperatures above  150~MeV, also provides a strong phenomenological motivation.


The study of the equation of state of the quark-gluon plasma at leading order in $\alpha_s$ within the DMB formalism was carried out in \cite{Baratella:2024sax}.
Beyond  $O(\alpha_s)$   one encounters  complications that hinge on a beautiful physical interpretation. 
One  technical difficulty is due to the fact that DMB evaluates amplitudes  in the strict forward limit and one needs to be 
  careful in dealing with the $i\varepsilon$ regulator. How to consistently do this limit is discussed at length in \cite{Baratella:2024sax} in the context of the effective theory of long strings.
All in all, forward singularities are cured by taking the limit 
${\varepsilon \to 0}$ only at the very end of the computation. For example, $\lim_{\varepsilon \to 0}(G_0(E)-\bar G_0(E))T(E)=-2\pi i \delta(E-H_0)T(E)$ when $T(E)$ is a regular function -- below we will encounter examples where this is not true when $T(E)$ is evaluated in the forward limit.


Once forward singularities are cured, one is left at $O(\alpha_s^2)$ with a thermally averaged three-particle phase space integral which is divergent in the infrared (IR). The presence of IR singularities at this order in QCD is well known in the standard approach to Th-QFT  --  they show up at three-loops \cite{Arnold:1994ps} -- and   come from diagrams of the form
\be\label{IR_Th-QFT}
\begin{tikzpicture}[line width=1.1 pt, scale=.7, baseline=(current bounding box.center)]
  \draw[vector] (0,0) circle (1.3cm);
  \draw[fill=gray!20] (-1.3,.1) circle (.5cm);
  \draw[fill=gray!20] (1.3,.1) circle (.5cm);
  \node at (0,.9) {\scriptsize ${\bm k}$};
  \node at (0,.5) {\scriptsize $n=0$};
\end{tikzpicture}~\sim~ \alpha_s^2\, \beta^{-1} \int {\dd^3 k} \, \frac{{\rm tr}\big[{\bm \Pi}(0,{\bm k})^2\big]}{{\bm k}^4}
\ee
where ${\bm \Pi}_{\mu\nu}(k_0,\bm k)$ is the gluon one-loop thermal self-energy, $\beta$ is the inverse temperature and the trace is taken over spacetime indices. Diagrams contributing to the one-loop self-energy are hidden in the two blobs in the figure (such that, overall, the diagram has three loops). The complete thermal loop over $k$ splits into a sum over Matsubara modes with frequency $k_0=2\pi n \beta^{-1}$, of which only the $n=0$ contribution is shown in \eq{IR_Th-QFT} since no IR problem arises from the modes with $n\neq 0$.  
The standard procedure to deal with this type of IR divergence consists in performing a resummation over diagrams with an arbitrary number of blobs inserted along the ${\bm k}$ loop. Physically, this is interpreted by longitudinal modes acquiring a nonzero thermal mass, so their propagator is effectively $({\bm k}^2+m_L^2)^{-1}$, where $m_L^2=\lim_{{\bm k}\to 0}{\rm tr}\big[{\bm \Pi}(0,{\bm k})^2\big]$.

\bigskip

In this work we show how the IR problem is cured within the DMB formalism in the $\lambda\phi^4$ theory. 
This theory shares many qualitative features  with the IR divergences of the NLO QCD equation of state that we just reviewed.  
In particular \emph{(1)} three-loop diagrams of the form shown in the figure are IR divergent, and \emph{(2)} one-loop blobs give a thermal mass $m\sim \sqrt\lambda \,\beta^{-1}$ to the scalar. 
Therefore this problem serves as  a warm-up for the case of QCD.

The DMB formalism \reef{master} involves a trace over the vacuum S-matrix operator. It may thus seem puzzling how the thermal mass emerges, and how one should address within the DMB formalism the IR divergences appearing at three loops.
It is precisely the purpose of this work to resolve this conundrum, paving the way to extend the DMB formalism to the current state-of-the-art calculation  of the perturbative thermal QCD equation of state.

The rest of the paper is divided in five sections, plus the conclusions and appendiccs. In Section~\ref{LOandNLO} the pressure at $O(\lambda)$ and the IR finite contributions at $O(\lambda^2)$ are computed. In Section~\ref{sec:singular} the evaluation of   $O(\lambda^2)$ IR singular diagrams is carried out. Finally, Sections~\ref{sec:daisy} and \ref{sec:IR} deal respectively with the so-called `daisy' and `superdaisy' resummations, showing how they emerge in the DMB formalism.  It is interesting to draw connections between standard approach to Th-QFT and this formalism, Appendix~\ref{LOmatch} explains this link. 
Three  other appendices are included with further details of calculations presented in the main text.

\section{Infrared finite contributions at LO and NLO}
\label{LOandNLO}

From here on, the concrete model under scrutiny will be a theory of  spin zero particles in 3+1 dimensions that are governed by the interaction Lagrangian density ${\cal L}=-\frac\lambda{4!}\,\phi^4$.  Generalisation to $N$ flavours will be discussed  when needed.

We work in the limit $m\ll T$. It is well known that this limit requires the resummation of hard thermal loops, see e.g.~\cite{Quiros:1999jp} for a review. We shall see below how this resummation is enforced on us within the DMB formalism.
In this section we discuss the regular  (non IR divergent) contributions to the pressure at leading order (LO) and next-to-leading order (NLO) in the coupling $\lambda$.

Before embarking into the computation, we remark two more important features of \reef{master}.
%
%
%
First, the suffix $c$ in ${\rm Tr}_c$ means that one has to consider only those contributions to \eq{master} that have a unique overall $\delta^{(d)}(0)$, where $d$ denotes the number of space dimensions. This guarantees that the free energy is extensive in the volume of the system, {\it i.e.} homogeneous of degree one in the volume.  The factor of proportionality, that is the free energy density, is equal to minus the pressure, {\it i.e.} $F(\beta,L)=-p(\beta)L^d$. Therefore \eq{master} provides the equation of state of the system. Second, we recall that \eq{master} is valid in any space dimension $d$ and we will find it convenient to  regulate the QFT via dimensional continuation.

\subsection{Free energy at LO}
\label{LOp}

\begin{figure}[t] 
   \centering
   \begin{tikzpicture}[line width=1.1 pt, scale=.7, baseline=(current bounding box.center)]
   
	\draw[red] (4,0)--(4.5,2.5);
	\draw[blue] (5,0)--(4.5,2.5);
	
	\draw[red](4,5)--(2,0);
		\draw[white,line width=3](2,5) to[bend right=15] (6,0);
	\draw[red](2,5) to[bend right=15] (6,0);
	\draw[red](6,5) to (4.5,2.5);
	
		\draw[white,line width=3] (5,5) to[bend right=15] (1,0);
	\draw[blue] (5,5) to[bend right=15] (1,0);
		\draw[white,line width=3](1,5)to(3,0);
	\draw[blue](1,5)to(3,0);
		\draw[white,line width=3](3,5)to[bend left=15](7,0);
	\draw[blue](3,5)to[bend left=15](7,0);
	\draw[blue](7,5)to(8,0);
		\draw[white,line width=3](8,5)to(4.5,2.5);
	\draw[blue](8,5)to(4.5,2.5);
	
	\node at (1,-.5) {1};
	\node at (2,-.5) {2};
	\node at (3,-.5) {3};
	\node at (4,-.5) {4};
	\node at (5,-.5) {5};
	\node at (6,-.5) {6};
	\node at (7,-.5) {7};
	\node at (8,-.5) {8};
	
	\node at (1,5.5) {1};
	\node at (2,5.5) {2};
	\node at (3,5.5) {3};
	\node at (4,5.5) {4};
	\node at (5,5.5) {5};
	\node at (6,5.5) {6};
	\node at (7,5.5) {7};
	\node at (8,5.5) {8};

	\filldraw (4.5,2.5) circle (5pt);
	
   \end{tikzpicture}
   
\caption{One among the $\binom 82 \,6!$ contractions that give a connected diagram in the $8\to 8$ sector; see \eq{OVnton}. The colouring of lines is immaterial and is only there to help following the two chains of contractions (described in the main text). The lengths of the two chains are 3:5 in this example.}
\label{fig:OV}
\end{figure}
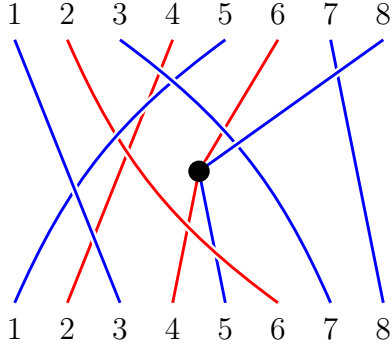

The leading $O(\lambda)$ contribution to \eq{master} comes from $n\to n$ forward amplitudes ($n\geq 2$) where two particles of the initial state scatter at tree level into two particles of the final state, while the remaining $n-2$ freely propagate. This contribution comes entirely from the first term in the expansion of $\ln S(E)$ in powers of $T(E)$ and the first term in the expansion of $T(E)$ in powers of $V$, defined in \eq{Lipp}. Using \eq{master} it is found, at $O(V)$,
\be\label{OV}
F=\int \dd a \int \dd E \,e^{-\beta E} \delta(E-E_a) \,V_{aa}\big|_c=\int \dd a \,e^{-\beta E_a} \,V_{aa}\big|_c\,,
\ee
where the notation $\int\dd a$ implies a sum-integral over the Fock basis. The restriction to trace-connected contributions is denoted with $|_c$. Here $V=\int\dd^3 x \phi^4$, so one needs to evaluate
\be\label{OVnton}
F_{n\to n}=\frac\lambda{4!}\int \dd^3 x\,\bigg(\frac1{n!}\,\prod_{i=1}^n\int \dd\Phi_i \,e^{-\beta |{\bm k}_i|}\bigg)\,\big\la 1,2,\ldots,n\big|\,\phi\,\phi\,\phi\,\phi\,\big|1,2,\ldots,n\big\ra_c\,.
\ee
The one-particle phase space measure is given by
\be
\dd\Phi_{i}=\frac{\dd^3k_i}{(2\pi)^32|{\bm k}_i|}\,.
\ee
Computing \reef{OVnton} 
is  a matter of doing Wick contractions. Particles 1 and 2 from the initial state can be fixed as the ones that interact on the $\phi^4$ vertex. There are $\binom n2$ ways to single out two special particles and $4\times3$ ways to choose which two $\phi$'s contract  with 1 and 2. Particles 1 and 2 in the final state can now either go to the vertex or contract with one among $3,4,\ldots,n$ in the initial state, and so on, until a chain of contractions is formed. 1 and 2 can form chains of different length $l$ and $(n-l)$, but overall the two chains need to include all particles in order for the diagram to be connected. The number of ways in which a $l:(n-l)$ pair of chains can be formed is $(n-2)!$, independent of $l$. Once the chain is completed, the remaining two `naked' final states need to contract with the leftover $\phi\,\phi$ in the vertex, and there are 2 inequivalent contractions. An example of a valid contraction is illustrated in Figure~\ref{fig:OV}.

A contraction among initial and final state particles gives a factor $\la i|j\ra=(2\pi)^32|{\bm k}_i|\delta^{(3)}({\bm k}_i-{\bm k}_j)$, while contractions with $\phi$ give factors of $e^{i{\bm p}\cdot {\bm x}}$ that eventually cancel because of momentum conservation at the vertex. Integrating over $\dd\Phi_3\ldots\dd\Phi_n$ with the Dirac deltas one finds
\be
F_{n\to n}=\frac1{2!}\,\lambda \,L^3\int \dd\Phi_1\,\dd\Phi_2 \sum_{l=1}^{n-1}e^{-\beta l|{\bm k}_1|}e^{-\beta (n-l)|{\bm k}_2|}\,,
\ee
where $L^3$ denotes the space volume. Doing the sum over $n\geq 2$ the exponentials reorganise themselves into a product of Bose-Einstein densities, and one immediately finds  $F=\frac12\lambda L^3\mu_{\rm th}^2$, where the thermally-averaged single-particle phase space measure is given by
\be\label{muth}
\mu_{\rm th}=\int \dd \Phi_{\bm k}\,n_B(|\bm k|)=\frac1{24\,\beta^2}\,,
\ee
with $n_B(|\bm k|)=(e^{\beta|\bm k|}-1)^{-1}$ the Bose-Einstein single particle density. As illustrated in \cite{Baratella:2024sax}, the free energy at leading order can be interpreted as a thermal average of the $2\to 2$ scattering amplitude at tree level, which here is simply equal to $\lambda$.

\subsection{Free energy at NLO: melon topology}\label{sec:melon}

Moving to $O(\lambda^2)$, one has to consider two types of contribution in the S-matrix formalism, coming from $2\to2$ and $3\to 3$ forward amplitudes, respectively at one loop and tree level. `Histories' that differ by the presence of an arbitrary number of freely propagating particles are resummed along the lines of the previous section.

At this order, given a forward scattering process, by connecting the final state legs to the initial state ones a topology is obtained that is either that of a `melon' or that of a `caterpillar', shown respectively in Figure~\ref{fig:melon}$(e)$ and \ref{fig:cater}$(f)$ (see also Appendix~\ref{LOmatch}). Different topologies translate into distinct kinematical properties of amplitudes, which in turn give a dramatically different singularity structure in the integrals that compute $F$. For these reasons the melon and caterpillar topologies are going to be the focus of separate sections.

\subsubsection{Free energy from $2\to2$ amplitude at one loop}

Due to the presence of UV divergences in loop amplitudes, we need to regulate the theory. We find it convenient to adopt dimensional regularisation, so we work in $d$ space dimensions and expand around $d=3$ at the end. The coupling constant $\lambda$ acquires a mass dimension $[\lambda]=3-d$. In order to keep track of dimensions, we set $\lambda=\hat\lambda m^{3-d}$.

\begin{figure}[t] 
   \centering
   
   \begin{tikzpicture}[line width=1.3 pt, scale=.8, baseline=(current bounding box.center)]
   
  	\draw (0,0) -- (0,2) ;
	\draw (1.5,0) -- (1.5,2) ;
	\draw (.75,1) circle (.75) ;
	\node at (0,-.4) {1} ;
	\node at (1.5,-.4) {2} ;
	\node at (0,2.4) {2} ;
	\node at (1.5,2.4) {1} ;
	\node at (2.8,1) {+} ;
	\node at (.75,-1.5) {$(a)$} ;
	
	\begin{scope}[xshift=4cm]
	\draw (0,0) -- (.75,.5) -- (1.5,0) ;
	\draw (0,2) -- (.75,1.5) -- (1.5,2) ;
	\draw (.75,1) circle (.5) ;
	\node at (0,-.4) {1} ;
	\node at (1.5,-.4) {2} ;
	\node at (0,2.4) {1} ;
	\node at (1.5,2.4) {2} ;
	\node at (2.8,1) {+} ;
	\node at (.75,-1.5) {$(b)$} ;
	\end{scope}
	
	\begin{scope}[xshift=8cm]
	\draw (0,0) -- (1.5,2) ;
	\draw (0,2) -- (1.,0) ;
	\draw (.5,2) -- (1.5,0) ;
	\node at (0,-.4) {1} ;
	\node at (1.,-.4) {2} ;
	\node at (1.5,-.4) {3} ;
	\node at (0,2.4) {3} ;
	\node at (0.5,2.4) {2} ;
	\node at (1.5,2.4) {1} ;
	\node at (2.8,1) {+} ;
	\node at (.75,-1.5) {$(c)$} ;
	\end{scope}
	
	\begin{scope}[xshift=12cm]
	\draw (0,0) -- (.75,.5) -- (1.5,0) ;
	\draw (0,2) -- (.75,1.5) -- (1.5,2) ;
	\draw (.75,2) -- (.75,0) ;
	\node at (0,-.4) {1} ;
	\node at (.75,-.4) {2} ;
	\node at (1.5,-.4) {3} ;
	\node at (0,2.4) {1} ;
	\node at (0.75,2.4) {2} ;
	\node at (1.5,2.4) {3} ;
	\node at (2.8,1) {$\equiv$} ;
	\node at (.75,-1.5) {$(d)$} ;
	\end{scope}
	
	\begin{scope}[xshift=16cm]
	\draw (.75,1) circle (1) ;
	\draw (.75,1) ellipse (.4 and .98) ;
	\node at (.75,-1.5) {$(e)$} ;
	\end{scope}
	
   \end{tikzpicture}

   \caption{$(a),(b)$: $2\to2$ one-loop amplitudes that contribute at $O(\lambda^2)$ to the free energy of the theory. $(c),(d)$: $3\to3$ tree-level amplitudes that contribute at the same order. $(e)$: When initial and final state particles with the same label are joined, all diagrams become topologically equivalent to the melon diagram.}
   \label{fig:melon}
\end{figure}
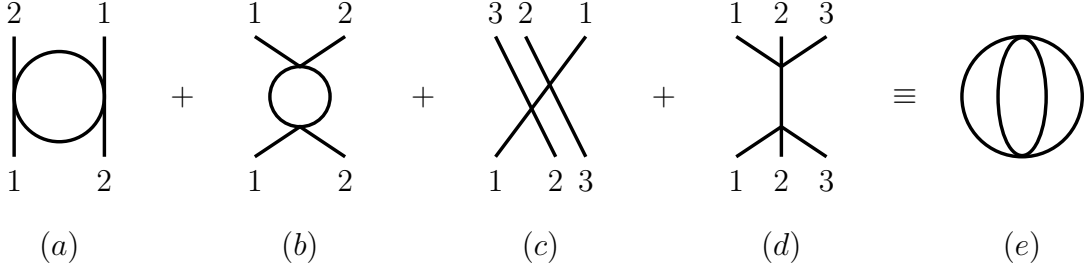

The $2\to2$ scattering amplitude at one loop is given by
\be\label{12to34loop}
{\la 34|T|12\ra^{(1)}}=\frac{\lambda^2}{2}\,(2\pi)^d\delta^{(d)}({\bm p}'-{\bm p})\,\bigg(B_d(-s)+{B}_d(-t)+{B}_d(-u)\bigg)\,,
\ee
where $B_d$ denotes the bubble integral in $d+1$ spacetime dimensions, and is equal to
\be
B_d(-s) =2\, (-s)^{\frac{d-3}2}\big(4\sqrt\pi\big)^{-d}\,\frac{\Gamma\big(\frac{3-d}2\big)\Gamma\big(\frac{d-1}2\big)}{\Gamma\big(\frac {d}2\big)} \, . 
\ee
Let us now consider the forward limit of \eq{12to34loop}, which amounts to taking $t\to0$. 
It can be seen that contributions proportional to either $B_d(-u)$ or $B_d(-s)$ belong to the melon topology: see diagrams $(a)$ and $(b)$ of Fig.~\ref{fig:melon}. These are also the ones which are regular in the forward limit. On the contrary, the term that is proportional to $B_d(-t)$, which is singular, belongs to the caterpillar topology and we deal with it in Section~\ref{sec:singular}.
Restricting to the melon (m) topology we find
\begin{align}\label{melon22}
\lim_{\rm forward}\,\la 34|T|12\ra^{(1)}\big|_{\rm m}=\frac{\lambda^2}2\,L^d \,\big(B_d(-s)+B_d(s)\big)\,.
\end{align}

We can now compute the free-energy associated to the contributions in \eq{melon22} by integrating the forward amplitude over the thermally averaged two-particle phase space, finding
\begin{align}\label{Fm2to2}
\frac{\beta^4}{L^3}\,&{\rm Re}\big[F_{(\ref{melon22})}\big]=\frac12\,\bigg(\frac{L}{m\beta^2}\bigg)^{\!d-3}\bigg(\prod_{i=1}^2\,\beta^{d-1}\int \dd\Phi_{d,{\bm k}_i}^{\rm th} \bigg) \,\hat\lambda^2\,B_d(sm^{-2}) \\
&=\frac{\hat\lambda^2}{32\pi^2}\bigg[\,\big(\mu_{{\rm th},d}\,\beta^{d-1}\big)^2\,\bigg(\frac{L}{m\beta^2}\bigg)^{\!d-3}\bigg(\frac 2{3-d}+2-\gamma_E+\ln4\pi\bigg)-\int\dd\Phi_{{\bm k}_1}^{\rm th}\dd\Phi_{{\bm k}_2}^{\rm th}\ln \frac s{m^2}\,\bigg]\,,\nonumber
\end{align}
where $\dd\Phi_{\bm k}^{\rm th}=\dd\Phi_{\bm k} n_B(|\bm k|)$ and $\mu_{{\rm th},d}=\beta^{1-d} \Gamma\big(\frac{d-1}2\big) \zeta(d-1)/(4 \,\pi^{\frac{d+1}2})$. 
The last integral, which is in $d=3$, can be easily performed by first integrating over the relative angle between $\bm k_1$ and $\bm k_2$, $\theta_{12}$, with $s=2|{\bm k}_1||{\bm k}_2|(1-\cos\theta_{12})$, to find
$
\int \dd\Phi_{{\bm k}_1}^{\rm th}\int\dd\Phi_{{\bm k}_2}^{\rm th}\,\ln \frac{s\,}{m^2}=\mu_{\rm th}^2\,(1-2\gamma_E-2\ln\frac{\beta m}{2}+\frac{12\zeta'(2)}{\pi^2})\,.
$
Putting everything together, one gets
\begin{align}\label{Fm2}
{{\rm Re}\big[F_{(\ref{melon22})}\big]}=\frac{\hat\lambda^2 \mu_{\rm th}^2L^3}{32\pi^2}\bigg(\frac 2{3-d}+1+3\gamma_E+3\ln\pi-2\ln\frac{L}{m^2\beta^3}-6\frac{\zeta'(2)}{\zeta(2)}\bigg)\,,
\end{align}
in perfect agreement with previous computations \cite{Frenkel:1992az,Arnold:1994ps}.
In \reef{Fm2} we took the real part because the full NLO contribution to the  free energy must be real, and thus imaginary parts are expected to cancel; see Appendix~\ref{appim}.

In order to get the free energy contribution of the melon topology in a theory with $N$ flavours one has to multiply \eq{Fm2} by $\frac13N(N+2)$.

\subsubsection{Free energy from tree-level $3\to3$ amplitude}\label{33mel}
At the same order the melon topology gives a contribution to the free energy through the following $3\to 3$ forward matrix element, depicted in Figure~\ref{fig:melon}$(c)$ and \ref{fig:melon}$(d)$,
\be\label{T123tree}
\frac{\la 123| T |123 \ra^{(0)}}{\lambda^2L^3}\bigg|_{\rm m}=\frac1{(p_1+p_2+p_3)^2}+\frac1{(-p_1+p_2+p_3)^2}+\frac1{(p_1-p_2+p_3)^2}+\frac1{(p_1+p_2-p_3)^2}\,.
\ee
The present contribution is both UV finite and regular in the forward limit, so it presents no conceptual difficulty. We find the same representation of the integral given in  Appendix E of \cite{Arnold:1994ps}, leading to 
\begin{align}\label{Fm3}
{\rm Re}\big[F_{(\ref{T123tree})}\big]=\frac{L^3T^4\lambda^2}{576(4\pi)^2}\bigg(\frac{\zeta'(-1)}{\zeta(-1)}-\frac{\zeta'(-3)}{\zeta(-3)}-\frac 7{15}\bigg)\,,
\end{align}
Taking the real part is equivalent here to performing the integral with the principal value prescription. 

\subsubsection{$0\to0$ and $1\to1$ contributions}

To conclude that \eq{Fm2} and \eq{Fm3} give the whole free energy in the melon sector, we need to make sure that no other diagram with  the same topology contributes. 
There are  other terms that naively could contribute but end up vanishing.

Here we discuss possible effects that come from 3-loop $0\to 0$ and 2-loop $1\to1$ amplitudes. In fact, nothing in \eq{master} seems to instruct us to exclude these terms. The $0\to0$ contribution, in very general terms (no restriction on topology nor number of loops), surely does add up to the energy content of the theory: it is the cosmological constant (notice that ${\rm Tr}_c$ means here that one has to consider connected diagrams with no external legs). However this contribution is independent of temperature. To sketch why formally this is the case, notice that $\int \dd E e^{-\beta E}\la 0|\delta(E-H_0)T|0\ra=\la 0|T|0\ra$, so temperature dependence goes away.

For what concerns $1\to1$ effects this argument does not apply, because $\int \dd E \,e^{-\beta E}\la {\bm p}|\delta(E-H_0)T|{\bm p}\ra=e^{-\beta |{\bm p}|}\la {\bm p}|T|{\bm p}\ra$  is temperature-dependent. One might argue that, in a physical renormalization scheme, given that ${\bm p}$ is on-shell, all such corrections can be made to vanish. 
Here  we show  directly that     the $1\to1$   contributions add up to zero.
We start by $1\to1$ effects at one loop, that is $O(\lambda)$. This means considering the tadpole diagram, which is zero for a massless scalar in dimensional regularisation, independent of the external momentum $p$. At $O(\lambda^2)$ there is the so-called sunset diagram, which has melon topology and is of the form \cite{Gehrmann:1999as}
\be\label{1to1}
\begin{tikzpicture}[line width=1.1 pt, scale=1, baseline=(current bounding box.center)]

		\draw (0,0) -- (2,0) ;
		\draw (1,0) circle (.5) ;
		\node at (-.25,-.05) {$p$};
		
		\node at (4,0.05) {$=~f(d)\,\big({-}p^2\big)^{d-2}\,.$};
	
 \end{tikzpicture}
\ee
This diagram is zero on shell, which proves that, up to imaginary terms, the temperature-dependent contribution to the free energy of the melon topology is given by the sum of \eq{Fm2} and \eq{Fm3}.

We dedicate Appendix~\ref{appim} to show that the imaginary parts of the diagrams with  melon topology cancel among themselves.

\section{Infrared singular contributions at NLO}\label{sec:singular}

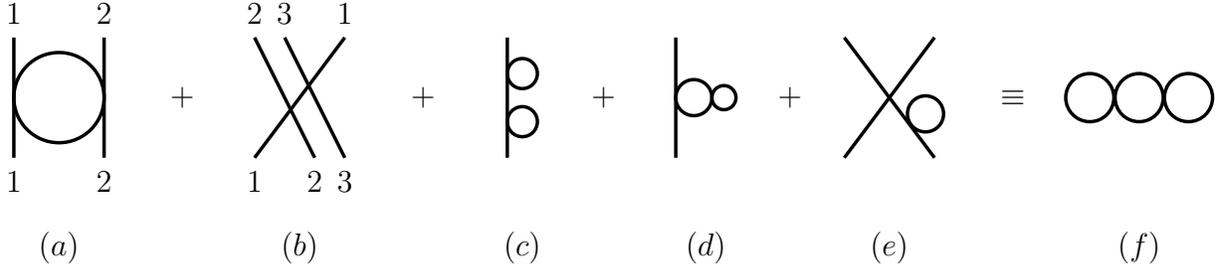
\begin{figure}[t] 
   \centering
   
   \begin{tikzpicture}[line width=1.3 pt, scale=.8, baseline=(current bounding box.center)]
   
  	\draw (0,0) -- (0,2) ;
	\draw (1.5,0) -- (1.5,2) ;
	\draw (.75,1) circle (.75) ;
	\node at (0,-.4) {1} ;
	\node at (1.5,-.4) {2} ;
	\node at (0,2.4) {1} ;
	\node at (1.5,2.4) {2} ;
	\node at (2.8,1) {+} ;
	\node at (.75,-1.5) {$(a)$} ;
	
	\begin{scope}[xshift=4cm]
	\draw (0,0) -- (1.5,2) ;
	\draw (0,2) -- (1.,0) ;
	\draw (.5,2) -- (1.5,0) ;
	\node at (0,-.4) {1} ;
	\node at (1.,-.4) {2} ;
	\node at (1.5,-.4) {3} ;
	\node at (0,2.4) {2} ;
	\node at (0.5,2.4) {3} ;
	\node at (1.5,2.4) {1} ;
	\node at (2.8,1) {+} ;
	\node at (.75,-1.5) {$(b)$} ;
	\end{scope}
	
	\begin{scope}[xshift=8.2cm]
	\draw (0,0) -- (0,2) ;	
	\draw (.25,.6) circle (.25) ;
	\draw (.25,1.4) circle (.25) ;
	\node at (.25,-1.5) {$(c)$} ;
	\node at (1.6,1) {+} ;
	\end{scope}
	
	\begin{scope}[xshift=11cm]
	\draw (0,0) -- (0,2) ;	
	\draw (.3,1) circle (.3) ;
	\draw (.8,1) circle (.2) ;
	\node at (.5,-1.5) {$(d)$} ;
	\node at (1.9,1) {+} ;
	\end{scope}
	
	\begin{scope}[xshift=13.8cm]
	\draw (0,0) -- (1.5,2) ;
	\draw (1.5,0) -- (0,2) ;
	\draw (1.35,.73) circle (.3) ;
	\node at (.75,-1.5) {$(e)$} ;
	\node at (2.8,1) {$\equiv$} ;
	\end{scope}
	
	\begin{scope}[xshift=18.7cm]
	\draw (0,1) circle (.4);
	\draw (-.82,1) circle (.4);
	\draw (.82,1) circle (.4);
	\node at (0,-1.5) {$(f)$} ;
	\end{scope}
	
   \end{tikzpicture}

   \caption{S-matrix elements that have caterpillar topology $(f)$
   and contribute to the temperature-dependent part of \eq{master}. All diagrams vanish in $d$ dimensions except $(b)$, which is divergent in the forward limit and has to be regulated by working with $E$-dependent $T$-matrix elements.}
   \label{fig:cater}
\end{figure}

In Section~\ref{sec:melon} we have considered the contribution to the free energy  that comes from $2\to 2$ amplitudes at one loop and $3\to 3$ amplitudes at tree level, restricting to diagrams that, once initial and final states are identified, belong to the melon topology.
Here we   complete the analysis of NLO effects by including the missing diagrams, those that are depicted in Figure~\ref{fig:cater}.

Like in the previous section, we work in generic number $d$ of space dimensions, and take the forward limit before the physical limit $d\to 3$. 
Note that dimensional regularization sets to zero   the   contribution from \eq{12to34loop}, i.e. the one proportional to the $t$-channel bubble $B_d(-t)\propto (-t)^\frac{d-3}2$, depicted in Figure~\ref{fig:cater}$(a)$. Indeed, taking the forward limit $t\to 0$ in $d$ dimensions gives zero for ${\rm Re}[d\,]>3$, and therefore for all $d$ by analytic continuation. Similarly, diagrams $(c)$, $(d)$ and $(e)$ of Figure~\ref{fig:cater} vanish in dimensional regularisation because they all contain one or two factors of the form $\int \frac{\dd^{d+1}p}{(p^2)^\alpha}=0$.

It only remains to consider diagram $(b)$. This time going to $d$ dimensions does not help, because the amplitude is proportional to
\be
\lim_{\rm forward} \frac1{(p_1+p_2-p_5)^2}
\ee
leading to a  $1/0$ singularity  when the identification $5\equiv 2$ is made (together with $4\equiv 1$ and $6\equiv 3$). As explained in \cite{Baratella:2024sax}, the cure to this sick behaviour lies in a more rigorous interpretation of \eq{master}. Until now we were able to define, at least operationally, $S(E)\equiv 1 - 2\pi i\delta(E-H_0)\, T$,
with the usual $T$-matrix of scattering theory. In other words, all the $E$-dependence went in the Dirac delta of energy conservation. However, the proper definition of the operator-valued distribution $S(E)$ appearing in \eq{master} is through the $E$-dependent operator $T(E)$, defined in \eq{Lipp}.

When we consider the matrix elements  $\la b|T(E)|a \ra$, we deal with $\mathbb{C}$-valued distributions in the kinematic variables defining the states $a$ and $b$, plus an additional energy variable $E$. If we keep $a$ and $b$ generic (i.e. $a\neq b$), it is possible to take the limit $E\to E_a$ and still have a well-defined distribution in the remaining variables. This is just the ordinary matrix element $T_{ba}$.

However, the evaluation of \eq{master} requires considering matrix elements in the forward limit, {\it i.e.} with $b=a$. In this limit ordinary $T$-matrix elements are not anymore necessarily well defined as distributions in the kinematic variables defining $a$, and may feature forward singularities. The extra energy variable $E$ turns out to be precisely the right ingredient to make forward matrix elements well defined, now as distributions in $a$ and $E$.  
To evaluate the off-shell matrix elements $\langle a |T(E)|a\rangle$ we will use Old Fashioned Perturbation Theory (OFPT) and dimensional regularisation. 

If a given contribution is safe from this kind of forward singularities, we can instead work directly with $\lim_{E\to E_a}\la a|T(E)|a\ra$, which is the ordinary scattering amplitude $T_{aa}$ that can be computed with the Feynman rules of Covariant Perturbation Theory \cite{Baratella:2024sax} -- this is what we did in the previous sections because forward divergences where not present.

\subsection{Free energy of caterpillar topology}\label{sec:misc}

In this Section we are going to see how, with the use of OFPT and the off-shell $T(E)$ operator, it is possible, following \eq{master}, to obtain a well-defined integrand for the seemingly intractable diagram \ref{fig:cater}$(b)$. 

In the OFPT way of organising the computation, working at $O(\lambda^2)$ means studying contributions to \eq{master} that have two insertions of the interaction Hamiltonian $V$. Similarly to what was done in Section~\ref{LOp}, we have to include processes with an arbitrary number of particles and the usual constraint of trace-connectedness. On top of this, the resulting diagram has to belong to the caterpillar topology.
Terms with two powers of $V$ come from two sources: either from the first term in the expansion of $\ln S(E)$, setting $T(E)=V\frac1{E-H_0+i\varepsilon}V$, or from the second term in the expansion of $\ln S(E)$, with $T(E)=V$.

For the formal developments of this Section it is convenient to introduce the following notation: when the resolvent $G(E)$ acts on an eigenstate of the free Hamiltonian it gives $G(E)|a\ra=G_a(E)|a\ra$, with $G_a(E)=(E-E_a+i\varepsilon)^{-1}$.

Expanding \eq{master} and restricting to $O(V^2)$ terms, one finds
\be\label{OV2}
F=-\frac1{2\pi i}\int \dd a\int \dd b \int \dd E \,e^{-\beta E} (G_a-\bar G_a ) V_{ab} \bigg( G_b -\frac12 (G_b-\bar G_b)\bigg) V_{ba}\,.
\ee
To obtain this result we have expanded $\ln S=(G-\bar G)T-\frac12 (G-\bar G)T(G-\bar G)T+O(T^3)$ and inserted the identity in the form of a projector $\int | b \ra\la b |$ to resolve the product of non-commuting operators. The $E$-dependence of $G_a$ and $G_b$ has been left implicit. Notice that the combination of $G_b$ and $\bar G_b$ that appears in \eq{OV2} is nothing but the real part of $G_b$, or $\frac12({G_b+\bar G_b})$.

\begin{figure}[t] 
   \centering
   
   \begin{tikzpicture}[line width=.8 pt, scale=.8, baseline=(current bounding box.center)]
		
		\draw (0,0) -- (5,0) -- (7.5,.5) -- (10,0);
		
		\draw (0,1) -- (2.5,1.5) -- (5,1) -- (7.5,.5) -- (10,1);
		
		\draw  (0,2) -- (2.5,1.5) -- (5,2) -- (10,2);
		
		\filldraw[white] (0,0) circle (4pt);
		\filldraw[white] (0,1) circle (4pt);
		\filldraw[white] (0,2) circle (4pt);
		\filldraw[white] (5,0) circle (4pt);
		\filldraw[white] (5,1) circle (4pt);
		\filldraw[white] (5,2) circle (4pt);
		\filldraw[white] (10,0) circle (4pt);
		\filldraw[white] (10,1) circle (4pt);
		\filldraw[white] (10,2) circle (4pt);
		
		\filldraw[white] (2.5,1.5) circle (6pt);
		\filldraw[white] (7.5,.5) circle (6pt);

		\filldraw (0,0) circle (1pt);
		\filldraw (0,1) circle (1pt);
		\filldraw (0,2) circle (1pt);
		\filldraw (5,0) circle (1pt);
		\filldraw (5,1) circle (1pt);
		\filldraw (5,2) circle (1pt);
		\filldraw (10,0) circle (1pt);
		\filldraw (10,1) circle (1pt);
		\filldraw (10,2) circle (1pt);
		
		\filldraw (2.5,1.5) circle (4pt);
		\filldraw (7.5,.5) circle (4pt);
		
		\node at (5,.6) {\scriptsize $1'$};
		\node at (5,-.4) {\scriptsize $3'$};
		\node at (5,1.6) {\scriptsize $2'$};
		
		\node at (0,.6) {\scriptsize 1};
		\node at (0,-.4) {\scriptsize 3};
		\node at (0,1.6) {\scriptsize 2};
		
		\node at (10,.6) {\scriptsize 1};
		\node at (10,-.4) {\scriptsize 3};
		\node at (10,1.6) {\scriptsize 2};
		
		\node at (2.5,.3) {\scriptsize $\la \,3\, |\,3'\,\ra$};
		\node[rotate=11.3099] at (3.8,2.05) {\scriptsize $\phi\, |\,2'\,\ra$};
		\begin{scope}[shift={(.6,-.6)}]\draw[rotate=11.3099,line width=0.3] (3.3,2.2) -- (3.3,2.35) -- (3.7,2.35) -- (3.7,2.2);\end{scope}
		
   \end{tikzpicture}

   \caption{Example of a Wick contraction that contributes to the caterpillar topology.}
   \label{fig:wick}
\end{figure}
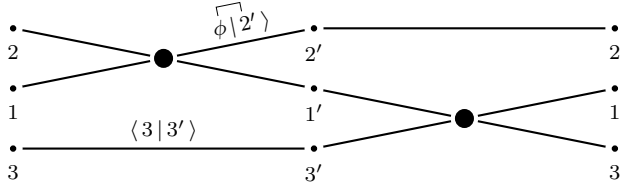
We next sum  over all diagrams with caterpillar topology. It is useful to start by looking at the concrete example of a contribution to \eq{OV2}, depicted in Figure~\ref{fig:wick}. This is an element of a restricted set of diagrams, called `seed diagrams', that have the property of having no `extra winding'. An extra winding is a line that flows in the diagram without ever encountering an interaction vertex. It is important to isolate the set of seed diagrams, because accounting for the effect of windings is then extremely easy: one needs simply to substitute $e^{-\beta|\bm k_i|}\to n_B(|\bm k_i|)$ for each external leg.

\begin{figure}[t] 
   \centering
   
   \begin{tikzpicture}[line width=1.1 pt, scale=.4, baseline=(current bounding box.center)]

   \node at (-9.5,3) {$A$};
   \node at (-4,3) {$B$};
   \node at (1.8,3) {$C$};
   \node at (8,3) {$D$};
   \node at (14.5,3) {$E$};

  \begin{scope}[xshift=-8cm]
  \draw (0,0) -- (0,6);
  \filldraw (0,2) circle (5pt);
  \filldraw (0,4) circle (5pt);
  
  \draw[red, thick, decorate, decoration={brace, amplitude=5pt}] 
    (-1.,6.5) -- (7.3,6.5);
  \node[red] at (3.2,8) {\small forward singular};
  
  \node[below, blue] at (0,-1.4) {$n_B$};
  \end{scope}
  
  \begin{scope}[xshift=-2.5cm]
  \draw (0,0) -- (0,6);
  \draw (1,0) -- (1,6);
  \filldraw (0,2) circle (5pt);
  \filldraw (1,4) circle (5pt);
  
  \node[below, blue] at (0.5,-1.1) {$n_B^2$};
  \end{scope}
  
  \begin{scope}[xshift=4cm]
  \draw (0,0) to [bend right=60] (0,4);
  \draw (0,2) -- (0,4);
  \draw (0,2) to [bend left=60](0,6);
  \filldraw (0,2) circle (5pt);
  \filldraw (0,4) circle (5pt);
  
  \node[below, blue] at (0,-1.4) {$n_B$};
  \end{scope}
  
  \begin{scope}[xshift=10.5cm]

  \draw (0,0) to [bend right=30] (0,4);
  \draw (1,0) to [bend right=45] (0,4);
  \draw (0,2) to [bend left=30](0,6);
  \draw (0,2) to [bend left=45](-1,6);
  \filldraw (0,2) circle (5pt);
  \filldraw (0,4) circle (5pt);
  
  \draw (5,0) to (5.5,2);
  \draw (6,0) to (5.5,2);
  \draw (5,6) to (5.5,4);
  \draw (6,6) to (5.5,4);
  \filldraw (5.5,2) circle (5pt);
  \filldraw (5.5,4) circle (5pt);
  \draw[blue, thick, decorate, decoration={brace, amplitude=5pt, mirror}] 
    (-1.5,-0.5) -- (6.5,-0.5);
  \node[below, blue] at (2.6,-1.1) {$n_B^2$};
  \end{scope}

  \draw[line width= .5] (-14,-5)--(21,-5);
  \draw[line width= .5] (1,-5)--(1,-15);

  	\begin{scope}[shift=({-10,-14})]
	\draw[gray!60] (2,0) -- (-1,6);
	\draw[gray!60] (1,0) -- (-2,6);
	\draw (0,0) -- (0,6);
	\filldraw (0,2) circle (5pt);
	\filldraw (0,4) circle (5pt);
	\node at (-2,7) {\footnotesize 2};
	\node at (-1,7) {\footnotesize3};
	\node at (0,7) {\footnotesize1};
	\node at (1,-1) {\footnotesize 2};
	\node at (2,-1) {\footnotesize3};
	\node at (0,-1) {\footnotesize1};
	
	\begin{scope}[xshift=6cm]
	\draw[gray!60] (1,0) -- (-2,6);
	\draw[gray!60] (2,0) -- (-1,6);
	\draw (-1,0) to [bend left=45] (0,4);
  	\draw (0,0) to [bend left=30] (0,4);
  	\draw (0,2) to [bend right=45](1,6);
  	\draw (0,2) to [bend right=30](0,6);
  	\filldraw (0,2) circle (5pt);
  	\filldraw (0,4) circle (5pt);
	\node at (2,-1) {\footnotesize 3};
	\node at (-1,-1) {\footnotesize1};
	\node at (0,-1) {\footnotesize$1'$};
	\node at (1,-1) {\footnotesize 2};
	\node at (1,7) {\footnotesize $1'$};
	\node at (-2,7) {\footnotesize2};
	\node at (0,7) {\footnotesize 1};
	\node at (-1,7) {\footnotesize3};
	\end{scope}
	
	\begin{scope}[xshift=15cm]
	
	\draw (0,0) -- (0,6);
	\draw (1,0) -- (1,6);
	\filldraw (0,2) circle (5pt);
	\filldraw (1,4) circle (5pt);
	\filldraw[red] (0,0) circle (3pt);
	\filldraw[red] (1,6) circle (3pt);
	\filldraw[blue] (0,6) circle (3pt);
	\filldraw[blue] (1,0) circle (3pt);
	
	\node at (3.9,3) {+};
	
	\end{scope}
	
	\begin{scope}[xshift=22cm]
	
	\draw (-.25,0) -- (-.25,6);
	\draw[blue] (.5,0)--(.5,6);
	\draw[blue] (1,0)--(1,6);
	\node[blue] at (2.5,3) {$\cdots$};
	\draw[blue] (3.8,0)--(3.8,6);
	\draw (3.8+.75,0) -- (3.8+.75,6);
	\draw[red] (5.3,0)--(5.3,6);
	\draw[red] (5.8,0)--(5.8,6);
	\node[red] at (7.3,3) {$\cdots$};
	\draw[red] (8.6,0)--(8.6,6);
	\filldraw (-.25,2) circle (5pt);
	\filldraw (3.8+.75,4) circle (5pt);
	\filldraw[red] (-.25,0) circle (3pt);
	\filldraw[red] (3.8+.75,6) circle (3pt);
	\filldraw[blue] (-.25,6) circle (3pt);
	\filldraw[blue] (3.8+.75,0) circle (3pt);
	
	\end{scope}

  	\end{scope}
   \end{tikzpicture}

   \caption{Upper row: seed diagrams with caterpillar topology. To account for windings one has to multiply by $n_B^p$, where $p$ is the number of external lines. Bottom-left: particles whose lines encounter only one interaction vertex are shown in grey for diagrams $A,D$. Bottom-right: taking diagram $B$ as an example, it is shown that independent windings should be considered for each external line. Upper and lower red dots are identified under the Trace, and similarly for the blue ones.}
   \label{fig:palu}
\end{figure}
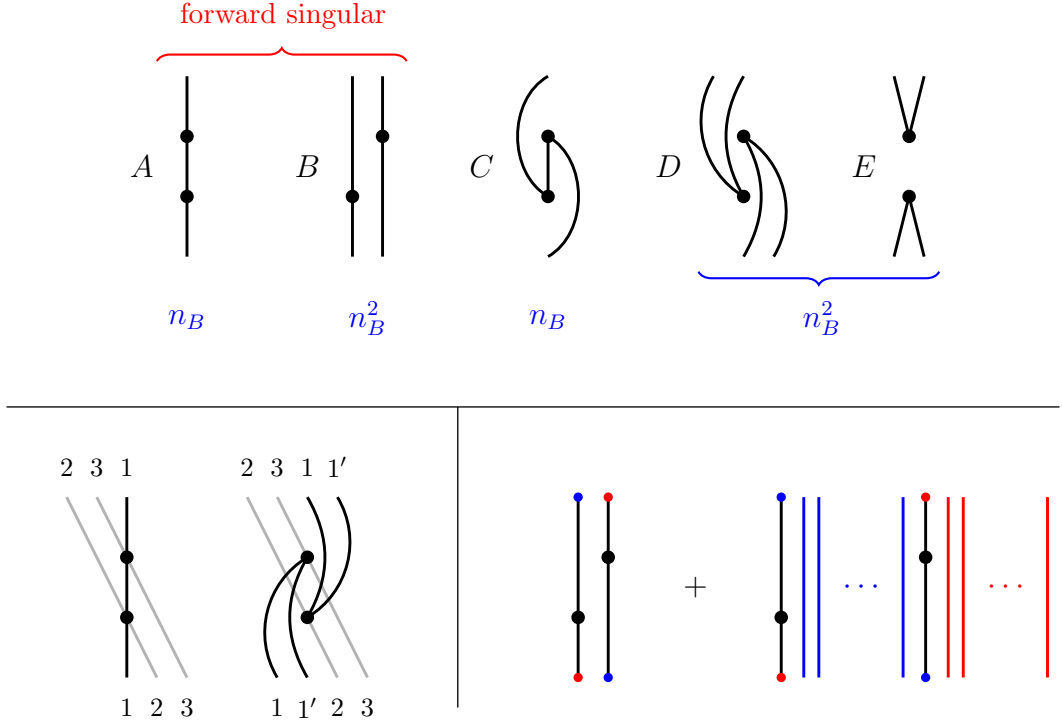

The seed diagram of Figure~\ref{fig:wick} has three external legs, and the operator product is resolved by the insertion of a three-particle state. Following \eq{OV2}, and multiplying by the number of equivalent Wick contractions, we find
\begin{align}
\hat F_{A}=-\frac{\lambda^2}{2\pi i}&\int \dd\Phi_1\,\dd\Phi_2\,\dd\Phi_3\int\dd\Phi_1'\,\dd\Phi_2'\,\dd\Phi_3'\int\dd E \,e^{-\beta E} (G_a-\bar G_a)\,\frac{G_b+\bar G_b}2 \int \dd^3x\int \dd^3 y \nonumber \\
&\times \bigg[\,\big\la   1\,2\,3\big|\,\phi\,\phi\,\phi\,\phi\,\big|1'\,2'\,3'  \big\ra\,\big\la 1'\,2'\,3'\big|\,\phi\,\phi\,\phi\,\phi\,\big|1\,2\,3\big\ra\,\bigg]_{\rm Fig.\,\ref{fig:wick}}\,,
\end{align}
where the suffix $A$ is an anticipation of Figure~\ref{fig:palu}, and the hat on $F$ is there to remind us that we need eventually to add windings. The Wick contraction gives
\begin{align}
\big[\,\la   a|V|b  \ra\la b|V|a\ra\,\big]_{\rm Fig.\,\ref{fig:wick}}&=\lambda^2\,\bigg(\prod_{i=2}^3(2\pi)^3 2|{\bm k}_i|\delta^{(3)}({\bm k}_i-{\bm k}_i')\bigg)\int \dd^3x\, e^{i{\bm x}\cdot({\bm k}_1-{\bm k}_1')}\int \dd^3y\, e^{-i{\bm y}\cdot({\bm k}_1-{\bm k}_1')}\nonumber \\
&=\lambda^2L^3\,\frac1{2|{\bm k}_1|}\prod_{i=1}^3\delta(\Phi_i-\Phi_i')\,.
\end{align}
States $|a\ra$ and $|b\ra$ stand for respectively $|123\ra$ and $|1'2'3'\ra$. Notice that the effect of $V_{ab}V_{ba}$ is primarily that of localising the $\int\dd b$ integration. The leftover is the expected volume factor, and a factor of $(2|{\bm k}_1|)^{-1}$ that comes from an incomplete reconstruction of $\delta(\Phi_1-\Phi_1')$. The integrals over $\dd\Phi_i'$, $\dd x$ and $\dd y$ basically drop out. This simplification will motivate a speedier approach in the spirit of Feynman rules, where the final result is directly readable from the diagram. For the moment, let us move on with this example step by step.

One crucial aspect of the contribution we are considering is that, after taking into account the various Dirac deltas, $b\equiv a$. This means the $E$ regulator is necessary to deal with this diagram. We use
\be\label{delta'}
(G_a-\bar G_a)\,\frac{G_a+\bar G_a}2=\frac12 (G_a^2-\bar G_a^2)=-\frac12\partial_E (G_a-\bar G_a)=\frac12 \,2\pi i\,\delta'(E-E_a)\,.
\ee
At this point one can perform the $\dd E$ integral with the $\delta'$, to get
\begin{align}
\hat F_A&=-\frac{\lambda^2L^3}2\int \dd\Phi_1\,\dd\Phi_2\,\dd\Phi_3\,\frac1{2|\bm k_1|}\int\dd E \,e^{-\beta E}\,\delta'(E-E_a)\nonumber \\&=-\frac{\beta\lambda^2L^3}2\bigg(\prod_{i=1}^3\int \dd\Phi_i \,e^{-\beta|\bm k_i|}\bigg)\frac1{2|\bm k_1|}\,.
\end{align}
The very final step consists in adding the effect of  windings, which gives
\be
F_{A}=-\frac\beta2\,L^3\,\lambda^2\,\mu_{\rm th}^2\,\int \dd\Phi_{\bm k} \,n_B(|{\bm k}|)\,\frac1{2|{\bm k}|}\,.
\ee
The end result is quite transparent and it is useful to comment on the various pieces. The dependence on $\lambda$ is obvious, and the proportionality to $L^3$ is also an omnipresent feature (due to the extensivity of $F$). A factor of $\mu_{\rm th}$ comes from each particle whose line encounters only one vertex, 2 and 3 in Figure~\ref{fig:wick}. The factor $-\frac12\beta$ is ultimately a consequence of being $b\equiv a$ for this diagram, which implies \eq{delta'} and an integration by part in $\dd E$, with the derivative hitting the Boltzmann factor $e^{-\beta E}$. The factor of $(2|\bm k|)^{-1}$ is associated to the internal line and is always present in OFPT, one for each propagator (see for example ``The old rule'' in \cite{Weinberg:1966jm}).

We now introduce a notation that we found very useful. 
The diagram of Figure~\ref{fig:wick} (which is time ordered in the horizontal direction) is simplified by stripping off the lines that correspond to particles that interact only once. Their effect is just a factor of $\mu_{\rm th}$ for each leg.  This is shown in Figure~\ref{fig:palu}. What remains is diagram $A$, where the dots represent the interaction vertices, which are vertically time-ordered, and the solid lines represent the state. How to restore the full diagram is shown in the bottom-left panel of Figure~\ref{fig:palu}.
Only particles that interact more than once can give rise to IR divergent propagators, thus this notation help us to focus the attention on those singularities.

There are four more diagrams that satisfy the three requirements of (i) being trace-connected, (ii) having the caterpillar topology and (iii) not having pure windings. They are shown in the upper panel of Figure~\ref{fig:palu}. We compute them in turn, starting from diagram $C$.

The first qualitative difference is that the intermediate state has five particles, that is $|b\ra=|1'1''1'''2'3'\ra$. Doing the Wick contractions for $\la   a|V|b  \ra\la b|V|a\ra$, it is found this time
\be
\lambda^2\,\delta(\Phi_1-\Phi_1')\delta(\Phi_2-\Phi_2')\delta(\Phi_3-\Phi_3')\delta(\Phi_1-\Phi_1''')\int \dd^3 x \,e^{i{\bm x}\cdot({\bm k}_1'+{\bm k}_1'')}\int \dd^3y\, e^{-i{\bm y}\cdot({\bm k}_1'+{\bm k}_1''')}\,.
\ee
The effect of the deltas is the same as before: one has just to remove the integration over $x,y,\Phi_i',\Phi_1'',\Phi_1'''$ and multiply by $L^3(2|{\bm k}_1|)^{-1}$. A subtle difference is the presence of $\delta^{(3)}({\bm k}_1+{\bm k}_1'')$, that imposes ${\bm k}_1''=-{\bm k}_1$. Diagrammatically this is reflected in the fact that one of the internal lines of $C$ flows `backwards'. However this leaves no trace because the integrand only depends on the absolute value of ${\bm k}_1''$. The crucial difference with respect to diagram $A$ is that $E_b$ is no longer equal to $E_a$, so we can set
\be
(G_a-\bar G_a)\,\frac{G_b+\bar G_b}2=-2\pi i \,\delta(E-E_a)\,\frac1{E_a-E_b}\,.
\ee
Staring at diagram $C$, the difference $E_a-E_b=-2|{\bm k}_1|$ can be understood immediately by counting the number of particles in the intermediate state and those in the asymptotic state. Notice, in this respect, that particles 2 and 3 just go through the diagram without changing their number, and the same can be said about pure windings, therefore they do not have any impact in the energy difference among internal and external states. This can be expressed as
\be\label{delta}
G_{a+\Delta}(E_{b+\Delta})=G_a(E_b)\,.
\ee
Performing the $\dd E$ integral with the Dirac delta and adding the effect of windings, one finds
\be
F_{C}=\,L^3\,\lambda^2\,\mu_{\rm th}^2\,\int \dd\Phi_{\bm k} \,n_B(|{\bm k}|)\,\frac1{-4|{\bm k}|^2}\,.
\ee
At the qualitative level, the main difference with $A$ is that the factor $-\frac12\beta$ is absent, and the energy difference denominator $(-2|\bm k|)^{-1}$ is there instead.

Diagrams $A$ and $C$ are the ones one would naively consider when trying to capture the contribution to the free energy of the caterpillar topology, as they correspond to the two time orderings that reproduce the covariant amplitude in Figure~\ref{fig:cater}$(b)$ \cite{Weinberg:1966jm}. However the two alone give the wrong result. Anticipating that $F_D+F_E=0$, we focus on the role of $B$, showing how it combines with $A$ to produce the correct result.

The first qualitative difference among $A$ and $B$ is that $|a\ra$ and $|b\ra$ have four particles instead of three. Quantitatively, $B$ is identical to a winding of $A$. However it is not the same object and should be counted as new, because it arises from a different Wick contraction. The fact that $A$ plus one winding and $B$ are different is reflected diagrammatically. In the first case, following the line of particle 1 one would encounter two interaction vertices at the first lap, and none at the second (the winding). In diagram $B$, instead, the line of 1 encounters one vertex in the first lap and the other in the second.

There are two independent windings for $B$, as shown in the bottom-right panel of Figure~\ref{fig:palu}. Therefore we need to multiply by two powers of $n_B(|\bm k_1|)$. All in all we find
\be
F_B=-\frac\beta2\,L^3\,\lambda^2\,\mu_{\rm th}^2\,\int \dd\Phi_{\bm k} \,n_B(|{\bm k}|)^2\,\frac1{2|{\bm k}|}\,.
\ee
Interestingly, $F_B$ can be combined with $A$ to produce a factor $-\beta(n_B+n_B^2)=n_B'$.

Accounting for $B$ gives a negligible change in the region $\beta|{\bm k}|\gg 1$ but, on the contrary, a drastic change in the opposite limit, where the integrand changes from $\beta{|{\bm k}|^{-1}}$ to ${|{\bm k}|^{-2}}$ in the infrared. The DMB formalism captures this statistical enhancement as a multiplication of exchange effects when the intermediate state is identical to the asymptotic one.

Let us conclude this analysis by considering diagrams $D$ and $E$. They are both regular in the forward limit, so the treatment is similar to $C$. The crucial point is that the energy difference $E_a-E_b$ is opposite for the two diagrams, while the rest is identical, and so the two cancel.

All in all we then have
\be\label{finalV2}
F_A+F_B+F_C+F_D+F_E=\frac{1}{4}\,L^3\,\lambda^2\,\mu_{\rm th}^2\int\dd\Phi_{\bm k}\left(\,\frac{n_B'}{|{\bm k}|}-\frac{n_B}{|{\bm k}|^2}\right),
\ee
The expression reproduces the known result.

The result in \eq{finalV2} has two major features. The first is that the integral factorises completely into single particle phase space integrations. This property is shared by the $O(\lambda)$ correction, but not by the diagrams with melon topology.
The second crucial feature is that the integration over $\dd\Phi_{{\bm k}}$ is divergent, due to the infrared region, as $\int \dd k/k^2$.

It is possible to regulate the IR divergences using dimensional regularisation. The integral over the modulus of ${\bm k}$ in \eq{finalV2} is continued in $d$ dimensions to
\be
\int_0^\infty \dd k \,k^{d-4}\big(k\, n_B'(k)-n_B(k)\big)=(2-d)\,\zeta(d-3)\,\Gamma(d-3)\,,
\ee
which is divergent for $d\to3$. Using $\beta^{d-1}\int \dd\Phi^{\rm th}_{d}= \Gamma\big(\frac{d-1}2\big) \zeta(d-1)/(4 \,\pi^{\frac{d+1}2})$ and  $\int \dd\Omega_{d-1}=\frac{2\pi^{d/2}}{\Gamma(d/2)}$ for the angular integration in $d$ dimensions, it is found that, up to finite terms,
\be
F(d)=\frac1{8\pi^2} \,L^3\, \hat\lambda^2\, \mu_{\rm th}^2\,\bigg(\frac1{d-3}+\frac{\gamma_E}2+\ln\frac{L}{m\beta^2}+\frac32\ln(4\pi)-48\ln A_G\bigg)\,,
\ee
where $A_G$ is Glaisher's constant.

\section{Infrared singular contributions at NNLO and Daisy resummation}\label{sec:daisy}

\begin{figure}[t] 
   \centering
   
   \begin{tikzpicture}[line width=1.1 pt, scale=.4, baseline=(current bounding box.center)]
   
	\draw[line width=1.8] (0,0)--(6,6);
	\draw[line width=.8] (3,0)--(0,6);
	\draw[line width=.8] (6,0)--(3,6);
	\filldraw (2,2) circle (5pt);
	\filldraw (4,4) circle (5pt);
	
	\begin{scope}[xshift=11cm]
		
		\draw[line width=1.8] (0,0)--(6,6);
		\draw[line width=.8] (3,0)--(0,6);
		\draw[line width=.8] (6,0)--(3,6);
		\draw[line width=.8] (4.5,0)--(1.5,6);
		\filldraw (2,2) circle (5pt);
		\filldraw (4,4) circle (5pt);
		\filldraw (3,3) circle (5pt);
		
	\end{scope}
	
	\begin{scope}[xshift=22cm]
		
		\draw[line width=1.8] (0,0)--(6,6);
		\draw[line width=.8] (3,0)--(0,6);
		\draw[line width=.8] (6,0)--(3,6);
		\draw[line width=.8] (4,0)--(1,6);
		\draw[line width=.8] (5,0)--(2,6);
		\filldraw (2,2) circle (5pt);
		\filldraw (4,4) circle (5pt);
		\filldraw (2+2/3,2+2/3) circle (5pt);
		\filldraw (2+4/3,2+4/3) circle (5pt);
		
	\end{scope}
	
	\node at (8.5,3) {+};
	\node at (8.5+11,3) {+};
	\node at (8.5+22,3) {+};
	\node at (8.5+25,3) {\large $\ldots$};
		
   \end{tikzpicture}

   \caption{Amplitudes belonging to the class that is resummed in Section~\ref{sec:daisy}. Due to the special topology, all propagators have the same momentum $\bm k$, which coincides with the momentum of one of the external particles. This is because each incoming thin line carries in a given vertex the same momentum that another outgoing thin line carries out from the same vertex. Therefore, a diagram at $O(\lambda^p)$ has $p-1$ singular propagators.  All momenta (internal and external) that are equal to $\bm k$ are drawn with a thick line.}
   \label{fig:comb}
\end{figure}
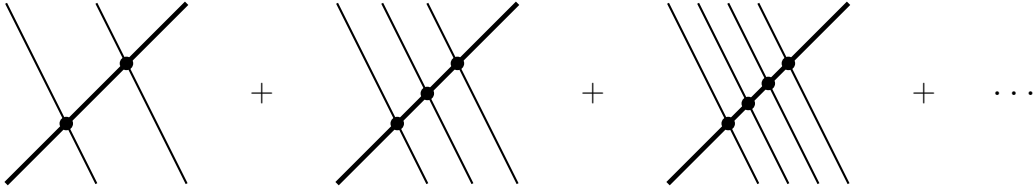

Proceeding with the analysis of higher order effects within the DMB formalism, we focus on the class of diagrams that is depicted in Figure~\ref{fig:comb}, which includes the diagram of Figure~\ref{fig:cater}$(b)$ as the leading element in a coupling expansion. The goal is twofold: on one side we want to show how to extend the method of the previous section to push the computation to $O(\lambda^3)$, and on the other side we aim at a formula that provides a resummation of all diagrams depicted in Figure~\ref{fig:comb}. At the end of the day it will be possible to identify the resummation with the well-known daisy resummation of Th-QFT \cite{Laine:2016hma}.

The class of diagrams we consider here has many features that allow to extend in an obvious way the analysis of Section~\ref{sec:misc} (see also the caption of Figure~\ref{fig:comb}). In order to show this, we explicitly compute with the DMB method the contribution to the free energy at $O(\lambda^3)$. Like at the previous order, we can focus on the special particle having momentum $\bm k$. All other particles give just a factor $\mu_{\rm th}^3$.

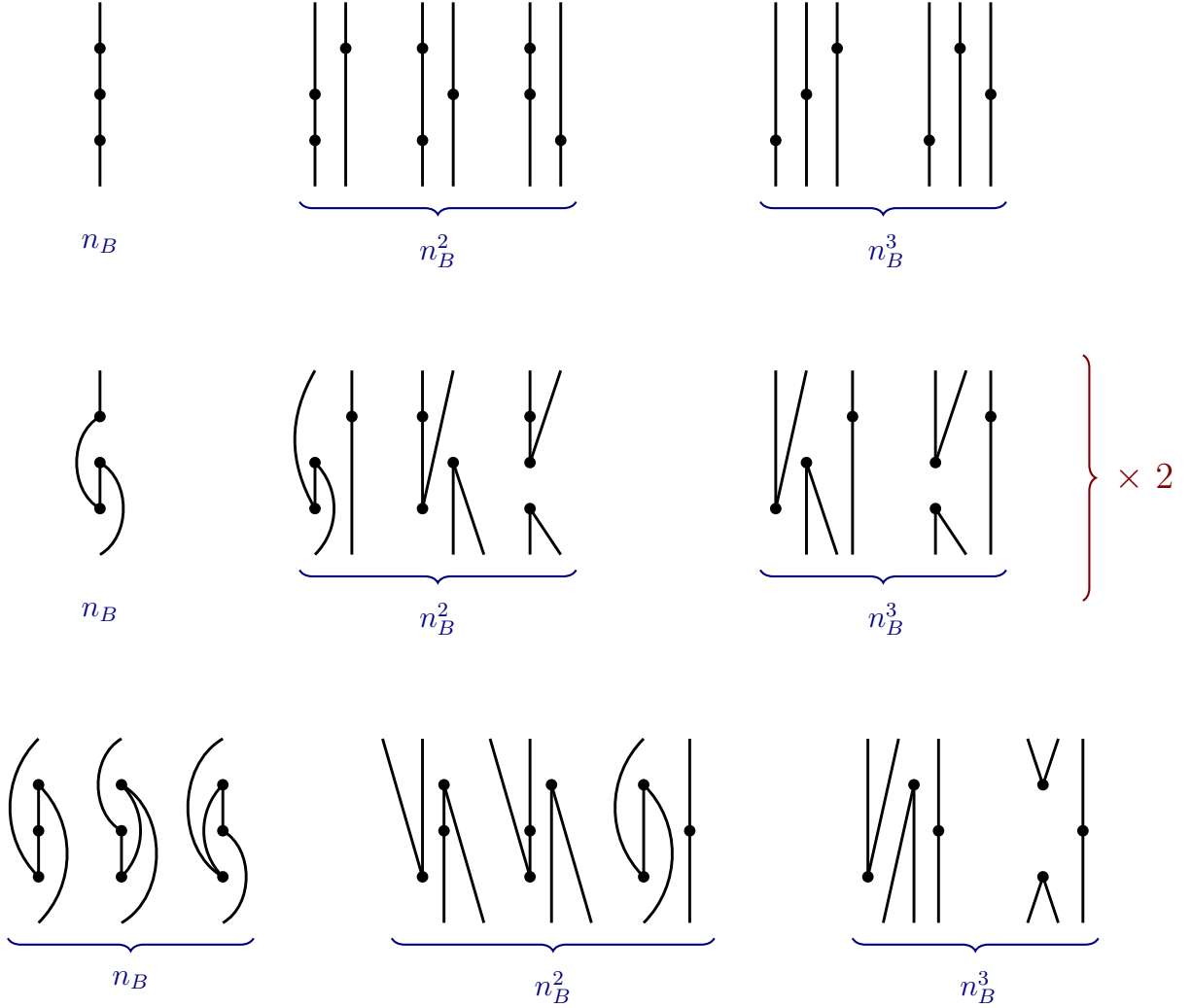
\begin{figure}[t] 
   \centering
   
\begin{tikzpicture}[line width=1.1 pt, scale=.42, baseline=(current bounding box.center)]
  
  \node[white] at (-12,3) {x};
  \begin{scope}[xshift=-7cm]
  \draw (0,0) -- (0,6);
  \filldraw (0,1.5) circle (4pt);
  \filldraw (0,3) circle (4pt);
  \filldraw (0,4.5) circle (4pt);
  
  \node[below, blue!50!black] at (0,-1.2) {$n_B$};
  \end{scope}
  
  \draw (0,0) -- (0,6);
  \draw (1,0) -- (1,6);
  \filldraw (0,1.5) circle (4pt);
  \filldraw (0,3) circle (4pt);
  \filldraw (1,4.5) circle (4pt);

  \draw (3.5,0) -- (3.5,6);
  \draw (4.5,0) -- (4.5,6);
  \filldraw (3.5,1.5) circle (4pt);
  \filldraw (4.5,3) circle (4pt);
  \filldraw (3.5,4.5) circle (4pt);

  \draw (7,0) -- (7,6);
  \draw (8,0) -- (8,6);
  \filldraw (8,1.5) circle (4pt);
  \filldraw (7,3) circle (4pt);
  \filldraw (7,4.5) circle (4pt);

  \draw[blue!50!black, thick, decorate, decoration={brace, amplitude=5pt, mirror}] 
    (-0.5,-0.5) -- (8.5,-0.5);
  \node[below, blue!50!black] at (4,-1.2) {$n_B^2$};

  \begin{scope}[xshift=15cm]
  \draw (0,0) -- (0,6);
  \draw (1,0) -- (1,6);
  \draw (2,0) -- (2,6);
  \filldraw (0,1.5) circle (4pt);
  \filldraw (1,3) circle (4pt);
  \filldraw (2,4.5) circle (4pt);

  \draw (5,0) -- (5,6);
  \draw (6,0) -- (6,6);
  \draw (7,0) -- (7,6);
  \filldraw (5,1.5) circle (4pt);
  \filldraw (7,3) circle (4pt);
  \filldraw (6,4.5) circle (4pt);

  \draw[blue!50!black, thick, decorate, decoration={brace, amplitude=5pt, mirror}] 
    (-0.5,-0.5) -- (7.5,-0.5);
  \node[below, blue!50!black] at (3.6,-1.2) {$n_B^3$};
  \end{scope}
  
  \begin{scope}[yshift=-12cm]
  
  \begin{scope}[xshift=-7cm]
  \draw (0,0) to [bend right=60] (0,3) to (0,1.5) to [bend left=60] (0,4.5) to (0,6);
  \filldraw (0,1.5) circle (4pt);
  \filldraw (0,3) circle (4pt);
  \filldraw (0,4.5) circle (4pt);
  
  \node[below, blue!50!black] at (0,-1.2) {$n_B$};
  \end{scope}
  
  \draw (0,0) to [bend right=45] (0,3) to (0,1.5) to [bend left=30] (0,6);
  \draw (1.2,0) to (1.2,6);
  \filldraw (0,1.5) circle (4pt);
  \filldraw (0,3) circle (4pt);
  \filldraw (1.2,4.5) circle (4pt);

  \draw (3.5,6) -- (3.5,1.5);
  \draw (3.5,1.5) -- (4.5,6);
  \draw (4.5,0) -- (4.5,3);
  \draw (4.5,3) -- (5.5,0);
  \filldraw (3.5,1.5) circle (4pt);
  \filldraw (4.5,3) circle (4pt);
  \filldraw (3.5,4.5) circle (4pt);

  \draw (7,6) -- (7,3);
  \draw (7,3) -- (8,6);
  \draw (7,0) -- (7,1.5);
  \draw (7,1.5) -- (8,0);
  \filldraw (7,1.5) circle (4pt);
  \filldraw (7,3) circle (4pt);
  \filldraw (7,4.5) circle (4pt);

  \draw[blue!50!black, thick, decorate, decoration={brace, amplitude=5pt, mirror}] 
    (-0.5,-0.5) -- (8.5,-0.5);
  \node[below, blue!50!black] at (4,-1.2) {$n_B^2$};
  
  \begin{scope}[xshift=15cm]
  \draw (0,6) -- (0,1.5);
  \draw (1,6) -- (0,1.5);
  \draw (1,0) -- (1,3);
  \draw (2,0) -- (1,3);
  \draw (2.5,0) -- (2.5,6);
  \filldraw (0,1.5) circle (4pt);
  \filldraw (1,3) circle (4pt);
  \filldraw (2.5,4.5) circle (4pt);

  \draw (5.2,6) -- (5.2,3);
  \draw (6.2,6) -- (5.2,3);
  \draw (5.2,0) -- (5.2,1.5);
  \draw (6.2,0) -- (5.2,1.5);
  \draw (7,0) -- (7,6);
  \filldraw (5.2,1.5) circle (4pt);
  \filldraw (5.2,3) circle (4pt);
  \filldraw (7,4.5) circle (4pt);

  \draw[blue!50!black, thick, decorate, decoration={brace, amplitude=5pt, mirror}] 
    (-0.5,-0.5) -- (7.5,-0.5);
  \node[below, blue!50!black] at (3.6,-1.2) {$n_B^3$};
  \draw[red!50!black, thick, decorate, decoration={brace, amplitude=5pt, mirror}] 
    (10,-1.5) -- (10,6.5);
  \node[red!50!black] at (12,2.5) {\large $\times ~2$};
  \end{scope}
  
  \end{scope}

    \begin{scope}[shift=({3cm,-24cm})]

	 \begin{scope}[xshift=-12cm]

  \draw (0,0) to [bend right=45] (0,4.5);
  \draw (0,4.5) to (0,1.5) to [bend left=45]  (0,6);
  \filldraw (0,1.5) circle (4pt);
  \filldraw (0,3) circle (4pt);
  \filldraw (0,4.5) circle (4pt);
  
  	\begin{scope}[xshift=-.3cm]
  \draw (3,0) to [bend right=60] (3,4.5);
  \draw (3,4.5) to [bend left=45] (3,1.5) to  (3,3) to [bend left=60] (3,6);
  \filldraw (3,1.5) circle (4pt);
  \filldraw (3,3) circle (4pt);
  \filldraw (3,4.5) circle (4pt);
  	\end{scope}
  
  \draw (6,6) to [bend right=60] (6,1.5);
  \draw (6,1.5) to [bend left=45] (6,4.5) to  (6,3) to [bend left=60] (6,0);
  \filldraw (6,1.5) circle (4pt);
  \filldraw (6,3) circle (4pt);
  \filldraw (6,4.5) circle (4pt);
  
  \draw[blue!50!black, thick, decorate, decoration={brace, amplitude=5pt, mirror}] 
    (-1,-0.5) -- (7.,-0.5);
  \node[below, blue!50!black] at (3,-1.2) {$n_B$};
  
  		\end{scope}
  
  	\begin{scope}[xshift=-.5cm]
  \draw (1,6) -- (1,1.5);
  \draw (1,1.5) -- (-.3,6);
  \draw (1.7,0) -- (1.7,4.5);
  \draw (1.7,4.5) -- (3,0);
  \filldraw (1,1.5) circle (4pt);
  \filldraw (1.7,3) circle (4pt);
  \filldraw (1.7,4.5) circle (4pt);
	\end{scope}
	
  \draw (4,6) -- (4,1.5);
  \draw (4,1.5) -- (2.7,6);
  \draw (4.7,0) -- (4.7,4.5);
  \draw (4.7,4.5) -- (6,0);
  \filldraw (4,1.5) circle (4pt);
  \filldraw (4,3) circle (4pt);
  \filldraw (4.7,4.5) circle (4pt);

  	\begin{scope}[xshift=.7cm]
  \draw (7,0) to [bend right=45] (7,4.5);
  \draw (7,4.5) to (7,1.5);
  \draw (7,1.5) to [bend left =45] (7,6);
  \draw (8.5,6) -- (8.5,0);
  \filldraw (7,1.5) circle (4pt);
  \filldraw (8.5,3) circle (4pt);
  \filldraw (7,4.5) circle (4pt);
  	\end{scope}

  \draw[blue!50!black, thick, decorate, decoration={brace, amplitude=5pt, mirror}] 
    (-0.5,-0.5) -- (10,-0.5);
  \node[below, blue!50!black] at (5.-.25,-1.2) {$n_B^2$};
  
  \begin{scope}[xshift=15cm]
  \draw (0,6) -- (0,1.5);
  \draw (1,6) -- (0,1.5);
  \draw (.5,0) -- (1.5,4.5);
  \draw (1.5,0) -- (1.5,4.5);
  \draw (2.3,0) -- (2.3,6);
  \filldraw (0,1.5) circle (4pt);
  \filldraw (2.3,3) circle (4pt);
  \filldraw (1.5,4.5) circle (4pt);

  \draw (5.2,6) -- (5.7,4.5);
  \draw (6.2,6) -- (5.7,4.5);
  \draw (5.2,0) -- (5.7,1.5);
  \draw (6.2,0) -- (5.7,1.5);
  \draw (7,0) -- (7,6);
  \filldraw (5.7,1.5) circle (4pt);
  \filldraw (5.7,4.5) circle (4pt);
  \filldraw (7,3) circle (4pt);

  \draw[blue!50!black, thick, decorate, decoration={brace, amplitude=5pt, mirror}] 
    (-0.5,-0.5) -- (7.5,-0.5);
  \node[below, blue!50!black] at (3.6,-1.2) {$n_B^3$};
  \end{scope}

  \end{scope}

\end{tikzpicture}

   \caption{seed diagrams relevant to the computation, with the DMB method, of the second term in Figure~\ref{fig:comb}.}
   \label{fig:paluV3}
\end{figure}

The first step is the identification of the class of seed diagrams, which are depicted in Figure~\ref{fig:paluV3}.
Like at $O(\lambda^2)$ the Dirac delta structure is such that all three space integrals and all phase space integrations over the intermediate particles go away, leaving a factor of $\lambda^3L^3(2|{\bm k}_1|)^{-2}$. Similarly, all the statistical factors at the denominator get exactly compensated by summing over the Wick contractions that correspond to an equivalent topology (cf. Section~\ref{sec:misc}).

After these first steps one has to study the different arrangements of the intermediate states. The crucial question to ask is if they have different or the same energy as the asymptotic state. Let us assume all three energies are generic, corresponding to a transition $a\to b\to c\to a$. From the expansion of $\ln S(E)$ up to $O(V^3)$ one gets the following product of resolvents
\be\label{OV3}
(G_a-\bar G_a)\left(G_c G_b-\frac12 G_c(G_b-\bar G_b)-\frac12(G_c-\bar G_c)G_b+\frac13 (G_c-\bar G_c)(G_b-\bar G_b)\right)\,.
\ee
Considering now the specific case of Figure~\ref{fig:paluV3}, one sees that the process is either of the form $a\to a\to a\to a$, like in the first row, or of the form $a\to b\to a\to a$ (second row), or finally of the form $a\to b\to b\to a$ (third row); in no case all three energies are distinct.
\begin{itemize}
\item In the first case, with all energies degenerate, we get the beautiful combination
\be
\frac13 \,(G_a^3-\bar G_a^3)=\frac16 \,\partial_E^2 (G_a-\bar G_a)=-\,\frac13\,\pi i\, \delta''(E-E_a)\,.
\ee
Notice that all terms need to delicately cooperate to get to this result.
\item In the second case we need to set $c\equiv a$, and it is natural to collect the first and third term of \eq{OV3}, to get
\be
\frac12 \,(G_a^2-\bar G_a^2)\, G_b+\underbrace{ (G_a-\bar G_a)\left(-\frac12 G_a+\frac 13 (G_a-\bar G_a)\right)(G_b-\bar G_b)}_{\text{cancels against the piece in the ellipses of \reef{tocone}}}= \pi i \,\delta'(E-E_a)\,G_b+\ldots
\label{toctwo}
\ee
where, in the right hand side, we have omitted the expression over the bracket. We show in Appendix~\ref{ic} that this term cancels against the ellipses in the right hand side of \reef{tocone}
after adding all the diagrams. 

\item For the configuration $a\to b\to b\to a$ we simply get
\be
\label{tocone}
(G_a-\bar G_a)\,G_b^2 -(G_a-\bar G_a)(G_b-\bar{G}_b)\left(\frac{2}{3}G_b+\frac{1}{3}\bar{G}_b\right) =-\,2\pi i \,\delta(E-E_a) \,G_b^2+\cdots
\ee
Terms hidden in the ellipses end up canceling and treated in Appendix~\ref{ic}. 

\end{itemize}
We are now ready to continue the computation. Using the previous results, the $\dd E$ integration can be carried out in general, and one finds
\begin{align}\label{dErulez}
-\frac1{2\pi i}\int \dd E \, e^{-\beta E} \begin{cases} 
       -\,\frac13\,\pi i\, \delta''(E-E_a)\\
       \pi i \,\delta'(E-E_a)\,G_b\\
       -\,2\pi i \,\delta(E-E_a) \,G_b^2
   \end{cases}=\,e^{-\beta E_a}\begin{cases} 
       \frac{\beta^2}6 \\
       -\frac12\big(\beta G_b(E_a)+G_b^2(E_a)\big)\\
       G_b^2(E_a)
   \end{cases}
\end{align}
where the term proportional to $G_b^2$ in the second line comes from $\partial_E$ hitting the $E$-dependent resolvent.

Adding one by one the seed diagrams of Figure~\ref{fig:paluV3} with the appropriate power of $n_B(|\bm k|)$, and using the rules of \eq{dErulez}, we get
\begin{align}\label{FV3}
F=&\,\,\lambda^3 L^3\mu_{\rm th}^3\int \dd \Phi_{\bm k} \,\frac1{4|\bm k|^2}\Bigg\{ \,\frac{\,\beta^2}6 \,\big(n_B+3n_B^2+2n_B^3\big) \nonumber -\beta\,\bigg[ \,n_B G_3(1)+n_B^2\big(2\,G_4(2)+G_0(2)\big)\\
&+n_B^3\big(G_5(3)+G_1(3)\big) \bigg]+\bigg[ \,n_B \big(3\,G_3^2(1)-G_3^2(1)\big)\\
&+n_B^2\big(3\,G_4^2(2)-2\,G_4^2(2)-G_0^2(2)\big)+n_B^3\big(G_5^2(3)+G_1^2(3)-G_5^2(3)-G_1^2(3)\big) \bigg] \,\,\Bigg\}\,,\nonumber
\end{align}
where we used the shorthand notation $G_m(p)=\frac1{p|{\bm k}|-m|{\bm k}|}$, with $p$ and $m$ counting respectively the asymptotic and internal lines. \eq{FV3} is organised according to the explicit powers of $\beta$ that appear in the integrand. Terms proportional to $\beta^2$ come from the first row of Figure~\ref{fig:paluV3}, and those proportional to $\beta$ from the second row. Notice that we have included a factor of 2 coming from diagrams that are the mirror image of the ones depicted in the second row. Terms with no power of $\beta$ come from the third row of Figure~\ref{fig:paluV3} when they have a positive sign, and from the second row when they have negative sign.

There are a lot of simplifications and cancellations happening in \eq{FV3}, the trace of an underlying structure which is invisible in this organisation of the computation. Using \eq{delta} and the fact that $G_p(m)=-G_m(p)$ we see that the last line cancels completely, and similarly the term in the second line proportional to $\beta n_B^3$. A partial cancellation takes place also in the terms of order $\beta n_B^2$ and $\beta^0n_B$. All in all we have
\be\label{finalV3}
F=\,\,\lambda^3 L^3\mu_{\rm th}^3\int \dd \Phi_{\bm k} \,\frac1{4|\bm k|^2}\bigg(\frac{1}6 n_B''+ n_B'G_3(1)+2n_BG_3(1)^2 \bigg)\,.
\ee
where we have left $G_3(1)=-\frac1{2|\bm k|}$ explicit to help the eye. It is notable that only combinations of powers of $n_B$ which are expressible as derivatives acting on $n_B$ are present. For example, we found the combination $\beta^2(n_B+3n_B^2+2n_B^3)=n_B''$.

The final result is largely decided by the first column of Figure~\ref{fig:paluV3}. However, to figure out the coefficients in front of $\frac1{(m+1)!}\partial^mn_B$, it is not enough to count the diagrams that have $m$ singular propagators, as one might naively think. This is because there are cancellations due to $\partial_E$ acting on the resolvent $G(E)$ instead of the Boltzmann factor $e^{-\beta E}$. In this example, we see that a 2 appears in front of $n_B$ as $2=3-1$, where 3 is the number of diagrams with no singular propagator, while the $-1$ comes from diagrams with one singular propagator, according to the second line of \eq{dErulez}.

\subsection{Emergence of thermal mass}

All in all, we have found
\begin{align}\label{daisyexp}
f=f_0+\frac{\lambda}{2}\mu^2_{\rm th}+\int \dd\Phi_{\bm k}\left((\lambda\mu_{\rm th})^2\,\frac{|\bm k|n'_B-n_B}{4|\bm k|^2}+(\lambda\mu_{\rm th})^3\,\frac{|\bm k|^2n''_B-3|\bm k|n'_B+3n_B}{24|\bm k|^4}\right)+O(\lambda^4)\,,
\end{align}
with $f=F/L^3$.
By comparing the NLO  \eq{finalV2} and NNLO \eq{finalV3}  we see a pattern emerging, 
\be 
f_{m}=\lambda^m \mu_{\rm th}^m \int {\dd\Phi_{\bm k}}{|\bm k|^{2-2m}}\left(
c_{m,m-1} |\bm k|^{m-1} n_B^{(m-1)}(|\bm k|)+ \cdots +
c_{m,1} |\bm k| n_B^\prime (|\bm k|)+ 
c_{m,0} \, n_B(|\bm k|) \right)\,.
\ee
Using the method explained above, the coefficients  $c_{m,i}$ can be systematically computed order by order.

\begin{figure}[t]
  \centering
  \includegraphics[width=.88\textwidth]{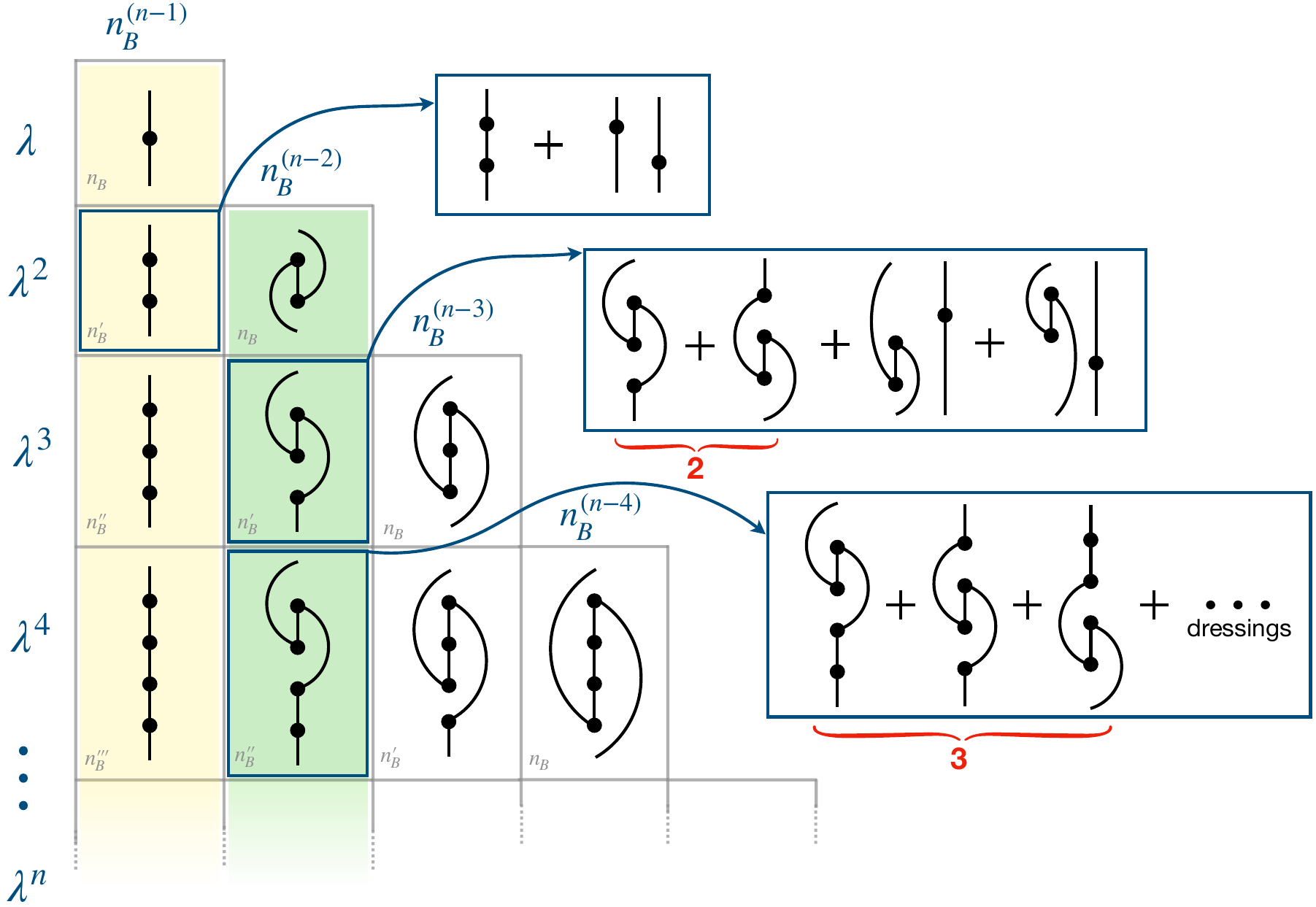} 
  \caption{Resummation of infrared singular diagrams.}
  \label{tabim}
\end{figure}

A different organising principle consists in identifying and summing up all the diagrams that contribute to the first two terms of the previous expression, at each order in $\lambda$, {\it i.e.}
\be
c_{m,m-1} |\bm k|^{m-1} n_B^{(m-1)}(|\bm k|) \quad  \text{and} \quad c_{m,m-2} |\bm k|^{m-2} n_B^{(m-2)}(|\bm k|)\,,
\ee
which amounts to computing the coefficients $c_{m,m-1}$ and $c_{m,m-2}$ for all $m$. The resummation of such terms  is all what is needed  to obtain an IR finite free-energy density at NLO -- {\it i.e.} to get rid of the IR divergence of the $\lambda^2$ term in \reef{daisyexp}.

Figure~\ref{tabim} depicts the two organising strategies. Diagrams are ordered vertically according to their power in $\lambda$, and horizontally according to the number of derivatives acting on $n_B$. 

Let us see how the resummation of the first column works. In each box we have drawn only one diagram -- leading in $e^{-\beta|{\bm k}|}$ -- that has to be properly dressed in order to reconstruct the full $n_B^{(n-1)}$ dependence. The dressing consists in adding diagrams that are product of lower order diagrams of the first column, such that overall the power in $\lambda$ is the correct one.
See for instance the upper blue box in Figure~\ref{tabim}, which contains diagrams $A$ and $B$ of Fig.~\ref{fig:palu}.
Another example to understand how the dressing of boxes in the first column works is the first row of Fig.~\ref{fig:paluV3}.
The dressing produces derivatives of the particle density, giving 
$
n_B^{(n-1)}
$
at order $\lambda^n$. For instance, the first blue box gives $n_B+n_B^2=-\beta n_B^\prime$.
Analogously $n_B+3n_B^2+2n_B^3=\beta^2 n_B^{\prime \prime }$, where the factors can be read from counting the diagrams in the first row of Fig.~\ref{fig:paluV3}, and so on. 
The evaluation of $n$-th diagram of the first column gives
$
a_n\equiv (\lambda \mu_\text{th})^n/n!\, \int d\Phi_{\bm k} n_B^{(n-1)}/(2|{\bm k}|)^{n-1}
$
which can be summed over $n\geq 2$ (the $O(\lambda)$ is IR finite by itself) to obtain
\be
\sum_{n=2}^\infty a_n=\beta^{-1}\int  \frac{d^3\bm k}{(2\pi)^3} \ln\bigg(1-e^{-\beta \left(|{\bm k}|+\tfrac{\lambda \mu_\text{th}}{2|{\bm k}|} \right)}\bigg) -a_{1}-a_0 \label{irone}\,.
\ee
Notice that the LO result $f_{\rm LO}=\frac12\lambda \mu_{\rm th}^2$ is equal to $\frac12 a_1$; the log resummation of \reef{irone} misses a symmetry factor of $\frac12$ that is only present in the $O(\lambda)$ diagram.

The resummation of diagrams in the second column proceeds analogously.
We first notice that the leading diagrams -- those that are represented in the boxes -- are all of the form
\be\label{blobnot}
 \begin{tikzpicture}[line width=1.1 pt, scale=.6, baseline=(current bounding box.center)]

		\draw (0,-2)--(0,2);
		\filldraw (0,-1.6) circle (3pt);
		\filldraw (0,1.6) circle (3pt);
		\filldraw (0,-.75) circle (3pt);
		\filldraw (0,.75) circle (3pt);
		\draw[fill=cyan] (0,0) circle (11pt);
		\node at (.5,-1) {$\vdots$};
		\node at (.5,1.3) {$\vdots$};
		
		\node at (4,0) {where};
		
		\draw (8,-1.) -- (8,1.) ;
		\draw[fill=cyan] (8,0) circle (11pt);
		\node at (9.5,0) {=};
		\draw (11,-1.1) to[bend left=60] (11,.4) -- (11,-.4) to[bend right=60] (11,1.1);
		\filldraw (11,.4) circle (3pt);
		\filldraw (11,-.4) circle (3pt);
		
		\node at (13,0) {.};

 \end{tikzpicture}
\ee
The dressing of the leading diagrams consist here in adding products of \emph{one} diagram of the second column times a number of diagrams of the first column (such that the desired order in $\lambda$ is reached). 

In taking the product of diagrams, it is important to treat both dots and blobs as interaction vertices that have a well defined time ordering. For example, the last diagram of order $n_B^2$ in the third row of Figure~\ref{fig:paluV3} is not allowed because it spoils the structure of the blob depicted in \reef{blobnot}. 
After adding the dressing we find 
$
b_n\equiv -(\lambda \mu_\text{th})^n/(n-2)!\, \int d\Phi_{\bm k} n_B^{(n-2)}/(2|{\bm k}|)^{n}
$. The combinatoric factor is easily obtained by noticing that the blob can be placed in $n-1$ positions. 
The sum gives
\be
\sum_{n=2}^\infty b_n = -\frac{\lambda^2\mu_\text{th}^2}{4} \int d\Phi_{\bm k} \frac{n_B(|{\bm k}|+  \tfrac{\lambda \mu_\text{th}}{2|{\bm k}|} )}{|{\bm k}|^2}   \label{irtwo}
\ee
which is IR finite. 

Equations \reef{irone} and \reef{irtwo} are nothing but the emergence of the thermal mass!

Indeed, 
consider the  free energy of a gas of free bosons with mass $m_D$, which is given by the IR finite expression (see \cite{Baratella:2024sax} for a derivation in the spirit of DMB)
\be\label{daisyF}
F=L^3 \beta^{-1} \int \frac{\dd^3 k}{(2\pi)^3}\ln\left(1-e^{-\beta \sqrt{\bm k^2+m_D^2}}\right)\,.
\ee
The thermal Debye mass is given by  $m_{D}=\sqrt{\lambda\mu_{\rm th}}$.
In order to get equations \reef{irone} and \reef{irtwo} from the Debye mass resummation in \reef{daisyF},
 consider the formal trick of expanding   \reef{daisyF} about $\sqrt{\bm k^2+m_D^2}-|{\bm k}|-\tfrac{m_D^2}{2|{\bm k}|}\ll1$.
 The first two terms of the expansion are indeed given by   \reef{irone} and \reef{irtwo}.
 To get the full resummation obtained from the introduction of the  Debye mass  requires adding the rest of the diagrams in Fig.~\ref{tabim}.
 
We have shown what are the  resummations that are needed in order to obtain an IR-finite  NLO free energy in the DMB approach. 
In the next Section we will show how the mass resummation works at all orders in the limit of large $N$ in the  $O(N)$ model.

\section{Large $N$ behaviour and Superdaisy resummation}\label{sec:IR}

We have identified in the previous section an infinite set of forward amplitudes with the property of being IR safe, despite being each amplitude divergent if taken alone. This property is certainly a good reason for considering the set as a whole, and in fact it also has a transparent physical interpretation, condensed in \eq{daisyF}.

In this section we would like to outline how to resum a larger set of amplitudes, including the set of Section~\ref{sec:daisy}.
It is made of tree level amplitudes whose propagators are all singular (for any configuration of external momenta). At $O(\lambda^p)$, this means considering $p+1\to p+1$ forward amplitudes that have all $p-1$ propagators on shell. The condition is highly selective but leaves a non-empty set, because all diagrams of Figure~\ref{fig:comb} respect it. Other examples are given in Figure~\ref{fig:4to4}.

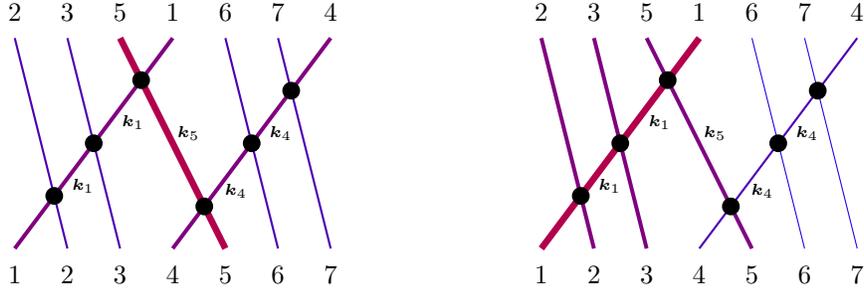
\begin{figure}[t] 
   \centering
   
   \begin{tikzpicture}[line width=1.1 pt, scale=.7, baseline=(current bounding box.center)]
   
  	\draw[blue!50!red,line width=1.5] (0,0)--(3,4);
	\draw[blue!70!red,line width=.8] (1,0)--(0,4);
	\draw[blue!70!red,line width=.8] (2,0)--(1,4);
	\draw[blue!50!red,line width=1.5] (3,0)--(6,4);
	\draw[blue!30!red,line width=2.5] (4,0)--(2,4);
	\draw[blue!70!red,line width=.8] (5,0)--(4,4);
	\draw[blue!70!red,line width=.8] (6,0)--(5,4);
	
	\filldraw (0.75,1) circle (4pt);
	\filldraw (1.5,2) circle (4pt);
	\filldraw (2.4,3.2) circle (4pt);
	\filldraw (3.6,.8) circle (4pt);
	\filldraw (4.5,2) circle (4pt);
	\filldraw (5.25,3) circle (4pt);
	
	\node at (0,-.5) {\footnotesize 1};
	\node at (1,-.5) {\footnotesize 2};
	\node at (2,-.5) {\footnotesize 3};
	\node at (3,-.5) {\footnotesize 4};
	\node at (4,-.5) {\footnotesize 5};
	\node at (5,-.5) {\footnotesize 6};
	\node at (6,-.5) {\footnotesize 7};
	
	\node at (0,4.5) {\footnotesize 2};
	\node at (1,4.5) {\footnotesize 3};
	\node at (2,4.5) {\footnotesize 5};
	\node at (3,4.5) {\footnotesize 1};
	\node at (4,4.5) {\footnotesize 6};
	\node at (5,4.5) {\footnotesize 7};
	\node at (6,4.5) {\footnotesize 4};
	
	\node at (2.25,2.4) {\tiny $\bm k_1$};
	\node at (1.3,1.2) {\tiny $\bm k_1$};
	\node at (5.05,2.2) {\tiny $\bm k_4$};
	\node at (4.2,1.1) {\tiny $\bm k_4$};
	\node at (3.3,2.2) {\tiny $\bm k_5$};
	
	\begin{scope}[xshift=10cm]
		
	\draw[blue!30!red,line width=2.5] (0,0)--(3,4);
	\draw[blue!50!red,line width=1.5] (1,0)--(0,4);
	\draw[blue!50!red,line width=1.5] (2,0)--(1,4);
	\draw[blue!70!red,line width=.8] (3,0)--(6,4);
	\draw[blue!50!red,line width=1.5] (4,0)--(2,4);
	\draw[blue!90!red,line width=.4] (5,0)--(4,4);
	\draw[blue!90!red,line width=.4] (6,0)--(5,4);
	
	\filldraw (0.75,1) circle (4pt);
	\filldraw (1.5,2) circle (4pt);
	\filldraw (2.4,3.2) circle (4pt);
	\filldraw (3.6,.8) circle (4pt);
	\filldraw (4.5,2) circle (4pt);
	\filldraw (5.25,3) circle (4pt);
	
	\node at (0,-.5) {\footnotesize 1};
	\node at (1,-.5) {\footnotesize 2};
	\node at (2,-.5) {\footnotesize 3};
	\node at (3,-.5) {\footnotesize 4};
	\node at (4,-.5) {\footnotesize 5};
	\node at (5,-.5) {\footnotesize 6};
	\node at (6,-.5) {\footnotesize 7};
	
	\node at (0,4.5) {\footnotesize 2};
	\node at (1,4.5) {\footnotesize 3};
	\node at (2,4.5) {\footnotesize 5};
	\node at (3,4.5) {\footnotesize 1};
	\node at (4,4.5) {\footnotesize 6};
	\node at (5,4.5) {\footnotesize 7};
	\node at (6,4.5) {\footnotesize 4};
	
	\node at (2.25,2.4) {\tiny $\bm k_1$};
	\node at (1.3,1.2) {\tiny $\bm k_1$};
	\node at (5.05,2.2) {\tiny $\bm k_4$};
	\node at (4.2,1.1) {\tiny $\bm k_4$};
	\node at (3.3,2.2) {\tiny $\bm k_5$};
	
	\end{scope}
	
   \end{tikzpicture}

   \caption{Alternative interpretations of the same $7\to7$ process. In the leftmost diagram, which arises at the second recursive level of \eq{ansatz}, $\bm k_5$ is treated as special. The rightmost diagram arises at the third recursion, with $\bm k_1$ treated as special. Using \eq{ansatz} implies an overcounting, because the above diagrams, which emerge as distinct contributions to $G$, correspond to exactly the same history.}
   \label{fig:nton}
\end{figure}

An important property of these diagrams is that, in a given configuration of external momenta ${\bm k}_1,\ldots,{\bm k}_n$, each internal momentum is exactly equal to one and only one ${\bm k}_i$. Furthermore, each `${\bm k}_i$ line' can be followed continuously from the initial to the final state, with no interruption. This property justifies the intuition that all particles $1,\ldots,n$ just proceed straight from the $|\text{\emph{in}}\rangle$ to the $|\text{\emph{out}}\rangle$  state, occasionally interacting with each other. 

In a theory with $N$ flavours, this set is important because it comprises all leading diagrams in the limit $N\to \infty$. Indeed, it can be seen that their contribution to the free energy per flavour, {\it i.e.} $\frac FN$, goes like $\big(\lambda (N+2)\big)^{p}$. This has to be contrasted with the melon topology, which defects a power of $N+2$.

We  provide here the  derivation of a formula for the sum of the whole set. This class of diagrams, called `superdaisy' or `foam' in the Th-QFT context, also has a long history; see for example \cite{Dolan:1973qd,Drummond:1997cw}. For a bottom-up resummation, using DMB, of the subclass of diagrams with the strongest forward divergence, see \cite{Baratella:2025tyb}.

In the previous section we saw  that diagrams with the structure of \eq{finalV2} and \eq{finalV3} can be interpreted, thanks to \eq{daisyF}, as providing a thermal mass to the gas constituents. The on-shell method gives a picture of the emergence of a thermal mass as due to consecutive forward scatterings of a given particle, that is single out, with a number of other particles of the thermal bath.
Shifting the attention from the  particle that we singled out to the thermal cloud, it is natural to ask what should be the parameters that control their distribution. In particular, we should expect that  we need to include a thermal mass also for particles in the bath.  Doing so would mean, in the  DMB approach, including diagrams where each particle of the thermal cloud interacts with a number of other particles that constitute its own thermal cloud (see Figure~\ref{fig:nton}). It is clear that this process is never-ending.

The thermal mass squared $m_D^2$ is given, at the lowest order in this recursive scheme, by the product of the coupling $\lambda$ and $\mu_{\rm th}$, the thermal phase space integral in three space dimensions given in \reef{muth}. Promoting everywhere in \reef{muth} and \reef{daisyF} $|{\bm k}|\to \sqrt{{\bm k}^2+m_D^2}$, we obtain an improved (recursive) definition of the Debye mass, along with an ansatz $G$ for the resulting free energy, that takes the same form \eq{daisyF}, but this time with a different meaning of $m_D$
\begin{align}
m^2_D&=\lambda \int \frac{\dd^3 k}{(2\pi)^3\,2\sqrt{{\bm k}^2+m_D^2}}\, n_B\big(\sqrt{{\bm k}^2+m_D^2}\big)\,, \label{recursM}\\
G&=L^3T\int \frac{\dd^3 k}{(2\pi)^3} \ln \left( 1- e^{-\beta\sqrt{\bm k^2+m_D^2}}\right)\,. \label{ansatz}
\end{align}
The recursive interpretation of diagrams like those in Figure~\ref{fig:nton} implies however an overcounting, and the expression $G$ for the free energy results incorrect. It is easy to understand where the overcounting comes from by comparing the two diagrams of Figure~\ref{fig:nton}. The one on the left is \emph{interpreted} as arising at the second recursive layer: the special momentum is $\bm k_5$, which interacts with $\bm k_1$ and $\bm k_4$ (first recursion), which in turn interact with respectively $\bm k_2,\bm k_3$ and $\bm k_6,\bm k_7$ (second recursion).
However there are as many ways to generate the same diagram as the number of particles, 7 in the present case. An example is given by the other diagram of Figure~\ref{fig:nton}, where the same process is generated at the third recursive level starting from $\bm k_1$.
The recursive formula \eq{ansatz} produces both, so we need to correct for the overcounting because they are the same exact amplitude.

At $O(\lambda^n)$, with $n+1$ particles partaking the scattering, there are $n+1$ ways to produce the same diagram from the expansion of $G$. If the ansatz is expanded as
\be
G(\lambda)=\sum_{n=0}^\infty \lambda^n G_n
\ee
the corrected free energy should be
\be\label{corrected}
F=\sum_{n=0}^\infty \lambda^n \,\frac{G_n}{n+1}=\frac1\lambda \int_0^\lambda \dd \lambda' \,G(\lambda')\,,
\ee
where the last expression is just implementing the algebraic operation of `dividing by $n+1$' at $O(\lambda^n)$. The reason why we do not have to similarly correct \eq{daisyF} is that, for all diagrams that are being resummed, there is an intrinsic way to define which one is the special particle: it is the one that interacts multiple times in its history (all others interact just once). Only at $O(\lambda)$, with $n=2$ particles in the process, this condition does not single out a special particle; see the discussion below \eq{irone}.

We are going to provide a couple of conceptually independent checks of the validity of \eq{corrected}, before moving to the conclusions.

\subsection{Check I: analytic method}

The first check comes from the well-known large temperature expansion of \eq{recursM} and \eq{ansatz}. They give
\begin{align}
m_D^2&=\frac\lambda{24}-\frac{\lambda m_D}{8\pi}+O(\lambda m_D^2\ln m_D)\,, \\
GL^{-3}&=-\frac{\pi^2}{90}+\frac{m_D^2}{24}-\frac{m_D^3}{12\pi}+O(m_D^4\ln m_D)=-\frac{\pi^2}{90}+\frac{\lambda}{576}-\frac{5\lambda^\frac32}{1152\sqrt6\,\pi}+O(\lambda^2\ln\lambda)\,,\label{Gcheck}
\end{align}
where the first expression, which is recursive, has been used in the second equality of \eq{Gcheck}. As well-know, the naive expansion in $\lambda$ is broken by the resummation, since terms of $O(\lambda^\frac32)$ are produced. Using \eq{corrected}, we find
\be\label{Fsuper}
FL^{-3}=\frac1\lambda\int^\lambda GL^{-3}=-\frac{\pi^2}{90}+\frac{\lambda}{1152}-\frac{\lambda^{3/2}}{576\sqrt{6}\,\pi}+O(\lambda^2\ln\lambda)\,.
\ee
The coefficient of the term of $O(\lambda^\frac32)$ has been computed by a number of authors using different methods, and the fact that we reproduce it is a good test of the correctness of the reasoning.

\subsection{Check II: combinatoric method}

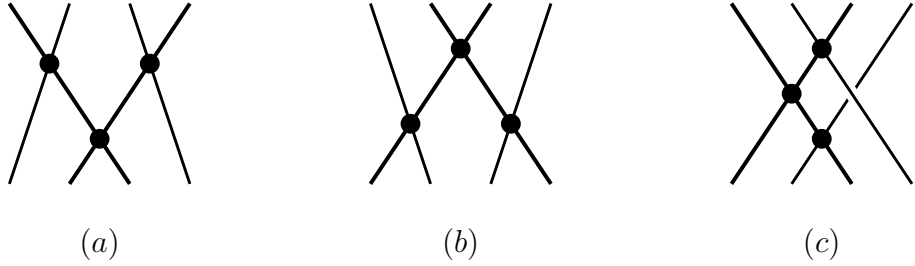
\begin{figure}[t] 
   \centering
   
   \begin{tikzpicture}[line width=1.1 pt, scale=.8, baseline=(current bounding box.center)]
   
  	\draw[line width=1.5] (2,0)--(0,3);
	\draw[line width=1.5] (1,0)--(3,3);
	\draw(0,0)--(1,3);
	\draw (3,0)--(2,3);
	
	\filldraw (1.5,.75) circle (4pt);
	\filldraw (.666,2) circle (4pt);
	\filldraw (3-.666,2) circle (4pt);
	
	
	\node at (1.5,-1.) {$(a)$};
	
	\begin{scope}[xshift=6cm]
	
		\draw[line width=1.5] (2,3)--(0,0);
		\draw[line width=1.5] (1,3)--(3,0);
		\draw(0,3)--(1,0);
		\draw (3,3)--(2,0);
	
		\filldraw (1.5,3-.75) circle (4pt);
		\filldraw (.666,1) circle (4pt);
		\filldraw (3-.666,1) circle (4pt);
		
		
		\node at (1.5,-1.) {$(b)$};
		
	\end{scope}
	
	\begin{scope}[xshift=12cm]
	
		\draw[line width=1.5] (0,0)--(2,3);
		\draw[line width=1.5] (2,0)--(0,3);
		\draw(1,0)--(3,3);
		\draw[white,line width=5] (3,0)--(1,3);
		\draw (3,0)--(1,3);
	
		\filldraw (1.5,2.25) circle (4pt);
		\filldraw (1.5,0.75) circle (4pt);
		\filldraw (1,1.5) circle (4pt);
		
		
		\node at (1.5,-1.) {$(c)$};
		
	\end{scope}
	
%
%
%
%
%

	\end{tikzpicture}

   \caption{$4\to4$ tree-level amplitudes that are maximally forward divergent and do not belong to the class of Figure~\ref{fig:comb}. They are leading in a large-$N$ expansion, as they give a contribution to $F$ proportional to $N(\lambda (N+2))^3$.}
   \label{fig:4to4}
\end{figure}

The second cross check consists in formally expanding \eq{corrected} in powers of $\lambda$, up to and including $O(\lambda^3)$, and check term by term if it produces the same result of the DMB method. Fully reproducing the $O(\lambda^3)$ effects will lead us to consider another set of forward divergent amplitudes, shown in Figure~\ref{fig:4to4}, whose computation we will sketch following the lessons of the previous sections. Expanding $G$ in powers of $\lambda$, we get
\begin{align}
GL^{-3}&=I_0+m_D^2I_1+m_D^4I_2+m_D^6I_3+O(\lambda^4)\,,
\end{align}
where $I_n$ are the $\dd\Phi_{\bm k}$ integrals that multiply $(\lambda\mu_{\rm th})^n$ in \eq{daisyexp}. To make the $\lambda$-dependence explicit we also need to expand $m_D$, which implicitly contains all orders in $\lambda$. This expansion needs to be carried out with care, because the expression is recursive and higher orders come from two sources, the Bose-Einstein density and the denominator $(\bm k^2+m_D^2)^{-\frac12}$. We get
\begin{align}
m_D^2&=\lambda I_1+2\lambda m_D^2 I_2+3\lambda m_D^4 I_3+O(\lambda^4)
=\lambda I_1+2\lambda^2I_1I_2+\lambda^3(4I_1I_2^2+3I_1^2I_3)+O(\lambda^4)\,.
\end{align}
In the final expression all dependence on $\lambda$ is made explicit. Plugging it into the expression for $G$, we get
\be
GL^{-3}=I_0+\lambda I_1^2+3\lambda^2I_1^2I_2+4\lambda^3\big(I_1^3I_3+2I_1^2I_2^2\big)+O(\lambda^4)\,.
\ee
Focusing on terms proportional to $(\lambda I_1)^n$, we notice that they are all off by a factor $n+1$ with respect to the correct result (the integral $I_1$ is the same as $\mu_{\rm th}$). This provides another set of data points to support the conjectured validity of \eq{Fsuper}. Performing the $\dd\lambda$ integral to adjust the combinatorial coefficients we get
\be\label{superDexp}
FL^{-3}=I_0+\frac12\lambda I_1^2+\lambda^2I_1^2I_2+\lambda^3\big(I_1^3I_3+2I_1^2I_2^2\big)+O(\lambda^4)\,.
\ee
It is interesting that, this time, the $O(\lambda)$ is reproduced with the right coefficient.
The prediction of the formula is that there will be another structure at $O(\lambda^3)$ beyond the one that was captured by \eq{daisyF}, proportional to $I_2^2$.

There are several ways to organise the OFPT computation of the new structure, which comes, loosely speaking, from the $E$-regulated version of the four amplitudes depicted in Figure \ref{fig:4to4}. The basic observation is that, out of the four lines with constant momentum, two lines encounter two vertices, and the other two only one. 

We can declare arbitrarily that the two lines which encounter two vertices have different colour, say red (r) and blue (b), while the remaining lines are, colloquially speaking, left black. The advantage of this is that we can now unambiguously define the three events that characterise the histories of Figure~\ref{fig:4to4}: `the blue line encounters the red line', `the red line encounters the black line' and `the blue line line encounters the other black line'. In OFPT, the six possible time orderings of the three events correspond to distinct processes one has to sum over.

Let us then fix a time ordering. Having done this, we can define a number of seed diagrams, on top of which we are going to add windings. From the combinatorial point of view, the problem is reduced to the square of Figure~\ref{fig:palu}: there are 5 paths the red line can follow, and similarly for the blue. Since the two choices of path are independent, there is a total of 25 seed diagrams. Following the notation of Figure~\ref{fig:palu}, each seed diagram can be naturally labelled with an ordered pair `rb', where each entry can take values $A,B,C,D,E$.

\begin{figure}[t] 
   \centering
   
   \begin{tikzpicture}[line width=1.1 pt, scale=.8, baseline=(current bounding box.center)]
  	
		\draw[blue!80!red] (0,0)--(0,2) to [bend left=-20] (.7,4);
		\draw[blue!20!red] (.7,0) to [bend left=-20] (0,2)--(0,4);
		
		\filldraw (0,1) circle (3pt);
		\filldraw (0,2) circle (3pt);
		\filldraw (0,3) circle (3pt);
		
		\node at (0.2,-.8) {\footnotesize$AA$};
		\node at (3.1,-.8) {\footnotesize$DC$};
		
		\draw[thick, decorate, decoration={brace, amplitude=5pt, mirror}] 
    		(-.8,-1.5) -- (4,-1.5);
 		\node[below] at (1.8,-2) {$t_{\rm b}<t_{\rm rb}<t_{\rm r}$};
		
		\begin{scope}[xshift=6cm]
		
		\draw[blue!80!red] (.6,0) to [bend right=45] (0,2)--(0,4);
		\draw[blue!20!red] (0.2,0) to [bend right=30] (0,2)--(0,1) to [bend left=30] (-.3,4);
		
		\filldraw (0,1) circle (3pt);
		\filldraw (0,2) circle (3pt);
		\filldraw (0,3) circle (3pt);
		
		\node at (0.1,-.8) {\footnotesize$CA$};
		\node at (2.8,-.8) {\footnotesize$BB$};
		
		\draw[thick, decorate, decoration={brace, amplitude=5pt, mirror}] 
    		(-.8,-1.5) -- (4,-1.5);
 		\node[below] at (1.8,-2) {$t_{\rm r}<t_{\rm rb}<t_{\rm b}$};
		
		\end{scope}
		\begin{scope}[xshift=12cm]
		
		\draw[blue!80!red] (0.25,0) --(0,1);
		\draw[blue!80!red] (0,1) -- (0.75,0);
		\draw[blue!20!red] (-.75,0)--(0,3) to [bend left=30] (0,1);
		\draw[white,line width=2.5]  (0,1) -- (-.75,4);
		\draw[blue!20!red]  (0,1) -- (-.75,4);
		\draw[white,line width=2.5] (0.25,4) to [bend left=30](0,2);
		\draw[white,line width=2.5] (0,2) to [bend left=-30] (0.75,4);
		\draw[blue!80!red] (0.25,4) to [bend left=30](0,2);
		\draw[blue!80!red] (0,2) to [bend left=-30] (0.75,4);
		
		\filldraw (0,1) circle (3pt);
		\filldraw (0,2) circle (3pt);
		\filldraw (0,3) circle (3pt);
		
		\node at (0,-.8) {\footnotesize$CE$};
		
		\draw[thick, decorate, decoration={brace, amplitude=5pt, mirror}] 
    		(-.8,-1.5) -- (.8,-1.5);
 		\node[below] at (0.1,-2) {$t_{\rm rb}<t_{\rm b}<t_{\rm r}$};
		
		\end{scope}
		\begin{scope}[xshift=3cm]
		
		\draw[blue!80!red] (-.3,0) to [bend left=30] (0,2) --(0,1) to [bend right=30] (.3,4);
		\draw[white,line width=2.5] (.3,0) to[bend right=20] (0,3);
		\draw[white,line width=2.5] (0,3) to [bend left=30] (.6,0);
		\draw[blue!20!red] (.3,0) to[bend right=20] (0,3);
		\draw[blue!20!red] (0,3) to [bend left=30] (.6,0);
		\draw[blue!20!red] (-.3,4) to [bend right=20] (0,2);
		\draw[blue!20!red] (0,2) to [bend left=30] (-.6,4);
		
		\filldraw (0,1) circle (3pt);
		\filldraw (0,2) circle (3pt);
		\filldraw (0,3) circle (3pt);
		
		\end{scope}
		\begin{scope}[xshift=9cm]
		
		\draw[blue!80!red] (-1.2,0) -- (.4,4);
		\draw[blue!80!red] (-.8,0) -- (.8,4);
		\draw[white,line width=2.5] (.4,0) -- (-1.2,4);
		\draw[white,line width=2.5] (.8,0) -- (-.8,4);
		\draw[blue!20!red] (.4,0) -- (-1.2,4);
		\draw[blue!20!red] (.8,0) -- (-.8,4);
		
		\filldraw (0,1) circle (3pt);
		\filldraw (0,2) circle (3pt);
		\filldraw (0,3) circle (3pt);
		
		\end{scope}

	\end{tikzpicture}

   \caption{Representative seed diagrams that enter in the evaluation of \eq{gridfinal}. Each diagram is uniquely determined by a choice of path for the red and the blue line, together with the specification of the ordering of events. The remaining lines are not drawn explicitly and can be reinserted following Figure~\ref{fig:palu}.}
   \label{fig:4to4palu}
\end{figure}
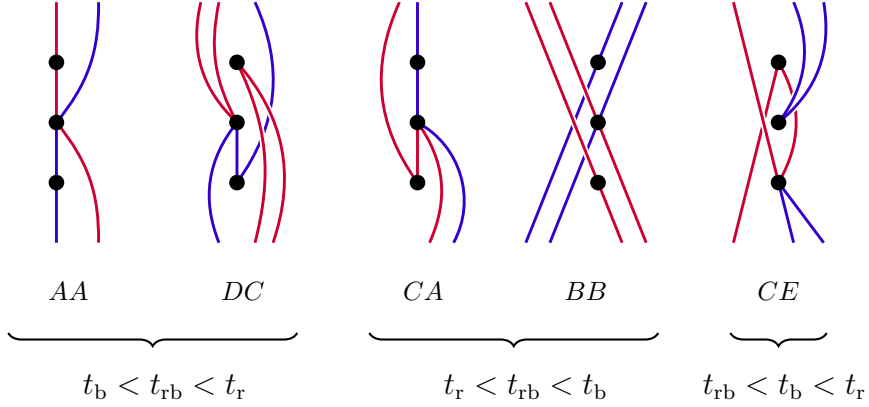

Some representative diagrams (5 out of 150) are depicted in Figure~\ref{fig:4to4palu}, where the `rb' labelling and the time ordering is specified. Their evaluation follows the rules outlined in the previous sections, with straightforward modifications.

For example, the leftmost diagram of Figure~\ref{fig:4to4palu} is equal to
\be
(AA)=\frac{\beta^2}6\,n_{\rm r}\,n_{\rm b}\,.
\ee
where we used \eq{dErulez} and multiplied by a Bose-Einstein density factor for each leg, the red and the blue, as they admit independent windings. Notice that we are using here an extremely synthetic notation: to obtain the free energy of a given diagram one has to integrate the skeleton expression according to
\be
F_{({\rm rb})}=L^3\,\lambda^3\,\mu_{\rm th}^2\int \dd\Phi_{\rm r}\,\dd\Phi_{\rm b}\,\frac1{4|\bm q_{\rm r}||\bm q_{\rm b}|}\,({\rm rb})\,,
\ee
where the factor of $\mu_{\rm th}^2$ comes from the two black lines.

The value of $AA$ is identical for all time orderings. $AA$ has two singular denominators in the forward limit and naturally groups together with $AB$, $BA$ and $BB$ (the latter is shown in Figure~\ref{fig:4to4palu}). Summing over all 6 time orderings gives
\be
\sum_{\substack{{\rm orderings} \\ i}}\bigg[(AA)+(AB)+(BA)+(BB)\bigg]_{ i}=n'_{\rm r}\,n'_{\rm b}\,.
\ee
An example of a regular contribution is given by diagram $DC$ in Figure~\ref{fig:4to4palu}. Its value, obtained from \eq{dErulez}, is
\be
(DC)_{t_{\rm b}<t_{\rm rb}<t_{\rm r}}=\frac1{-2|\bm q_{\rm r}|}\,\frac1{-2|\bm q_{\rm b}|}\,n_{\rm r}^2\,n_{\rm b}\,.
\ee
It is proportional to two powers of $n_{\rm r}$, because the red leg admits two independent windings.

An example where the importance of the ordering of events is clearly visible is $CA$. The central diagram of Figure~\ref{fig:4to4palu} is of $CA$ type and respects the ordering $t_{\rm r}<t_{\rm rb}<t_{\rm b}$. It corresponds to a contribution with one singular denominator. The ordering $t_{\rm r}<t_{\rm b}<t_{\rm rb}$ is instead regular in the forward limit. Following the rules of \eq{dErulez} one finds
\begin{align}
(CA)_{t_{\rm r}<t_{\rm rb}<t_{\rm b}}&=\left(\frac\beta{4|\bm q_{\rm r}|}-\frac1{8|\bm q_{\rm r}|^2}\right)n_{\rm r}\, n_{\rm b}\,,\\
(CA)_{t_{\rm r}<t_{\rm b}<t_{\rm rb}}&=\frac1{4|\bm q_{\rm r}|^2}\, n_{\rm r}\, n_{\rm b}\,.
\end{align}
Not all contributions can be written as a product of a function of $|\bm q_{\rm b}|$ and a function of $|\bm q_{\rm r}|$. This is shown by the last example
\be
(CE)_{t_{\rm rb}<t_{\rm b}<t_{\rm r}}=\frac1{-2|\bm q_{\rm r}|}\,\frac1{-2|\bm q_{\rm r}|+2|\bm q_{\rm b}|}\, n_{\rm r}\,n_{\rm b}^2\,,
\ee
whose diagram is the rightmost of Figure~\ref{fig:4to4palu}. The expected factorisation of the final result, which comes from summing over all seed diagrams for all possible time orderings, is therefore not obvious diagram by diagram.

The value of each individual seed diagram is given in Appendix~\ref{app:grid}. Their sum yields
\be\label{gridfinal}
L^{-3}\sum_{{\rm r,b}}F_{({\rm rb})}=\lambda^3\,\mu_{\rm th}^2\int \dd\Phi_{\rm r}\,\frac1{2|\bm q_{\rm r}|}\,\bigg( n_{\rm r}' - \frac{n_{\rm r}}{|\bm q_{\rm r}|}\bigg)\int\dd\Phi_{\rm b}\,\frac1{2|\bm q_{\rm b}|}\,\bigg( n_{\rm b}' - \frac{n_{\rm b}}{|\bm q_{\rm b}|}\bigg)=4\lambda^3I_1^2I_2^2\,,
\ee
which is off by a factor 2. The reason is that we are treating histories that differ by an interchange of r and b as distinct contributions to the free energy. But the colouring has been introduced artificially and we have to compensate for this: dividing \eq{gridfinal} by 2  we find perfect agreement with the prediction of \eq{superDexp}.

\section{Outlook}

In this work, we show how to handle the perturbative IR divergences in the DMB formalism that give rise to the Debye mass.
Moving forward, the very same mechanism applies at NLO in QCD, and it would be worthwhile to carry out the computation explicitly.

Further interesting  avenues include the generalisation of the  DMB approach to other observables besides the free energy~\cite{Jeon:1998zj,Baratella:2024sax}. 
Following similar steps that lead to \reef{master}, it is possible to express thermal $n$-point functions  in terms of the  S-matrix operator  and form-factors.

\subsubsection*{Acknowledgements}

We thank   Emanuele Gendy for early collaboration on the model considered here. We thank  Giovanni Villadoro  and Andrea Wulzer for useful discussions. 
JEM is supported by the European Research Council, grant agreement n. 101039756.
PB is supported by the Slovenian Research
Agency (research core funding No. P1-0035 and J1-4389). 

%

\appendix

\section{Matching to Th-QFT}\label{LOmatch}
It is natural to assign to each forward diagram, like the ones depicted in Figure \ref{fig:OV}, a simpler diagram which has the property of being connected and without external legs.
One just needs to join each leg of the final state to the one in the initial state that has the same label.  

Under the in-out identification, all diagrams that contribute to $F$ at $O(\lambda)$ are topologically equivalent to the only Th-QFT diagram that arises at the same order, which can be expressed as \cite{Gynther:2007bw}
\be\label{thO1}
  \begin{tikzpicture}[line width=1.1 pt, scale=.7, baseline=(current bounding box.center)]
  	\draw (.73,0) circle (.7) ;
	\draw (-.7,0) circle (.7) ;
\end{tikzpicture}~=~ \frac{\lambda}{8} \left(\beta^{-1}\sum_{m\in {\mathbb Z}}\int \frac{\dd^3 p}{(2\pi)^3}\frac{1}{{\bm p}^2+(2\pi m \beta^{-1})^2}\right)^2\,.
\ee
Thanks to the Poisson summation formula, it is possible to rewrite
\be\label{mtoell}
\sum_{m\in {\mathbb Z}}\frac{1}{{\bm p}^2+(2\pi m \beta^{-1})^2}=\beta\,\sum_{\ell\in{\mathbb Z}}\int_{-\infty}^{+\infty}\frac{\dd \omega}{2\pi}\frac{e^{- i\ell\beta\omega}}{{\bm p}^2+\omega^2}=\frac{1}{2|{\bm p}|}\bigg( 1+2\sum_{\ell>0}e^{-\ell\beta |{\bm p}|}\bigg)\,,
\ee
where, in the second step, the integral over $\omega$ has been evaluated by closing the contour in the upper or lower complex half-plane for respectively negative and positive $\ell$.

The first term can be recognised as the divergent zero-temperature tadpole integral, which can be consistently set to zero (e.g. with dimensional regularisation). What remains is ${|\bm p|}^{-1}n_B(|{\bm p}|)$. After plugging this into \eq{thO1}, we get as expected $\frac12\lambda \mu_{\rm th}^2$.
With this example in mind, let us comment on the relation among DMB and Th-QFT.
\begin{itemize}
\item There is only one diagram at $O(\lambda)$ in Th-QFT, corresponding to infinitely many histories from the amplitude perspective. However, the additional diagrams only differ from the leading one with $n=2$ by the addition of freely propagating particles. Therefore all the complexity lies in the single leading diagram proportional to a $2\to 2$ scattering amplitude.

\item Using \eq{mtoell} it is possible to decompose \eq{thO1} into pieces having definite $\ell_1,\ell_2>0$. Terms with $\ell_1+\ell_2=n$ correspond, on the DMB side, to all the contributions coming from $n\to n$ amplitudes.
For example, the diagram of Figure~\ref{fig:OV} comes from choosing $\ell_1=3,\ell_2=5$. One can imagine taking the diagram in \eq{thO1} and mapping it in a continuous way on a tube: $\ell_i$ counts the number of times leg $i$ is winded around the tube.

\item Th-QFT integrals are defined with Euclidean signature. In going from the `Matsubara basis' to the `winding basis' a contour integral in the energy variable $\omega$ naturally arises, which can be reduced to a sum over residues at the poles of the Euclidean propagators. Since poles are located at imaginary $\omega$, performing the contour integral via the residue theorem has the effect of localising the integration on Lorentzian points. At the same time, reducing the integral to a sum over propagator poles means that \emph{loop momenta are put on-shell}. The DMB formulation that is explored here directly captures the on-shell nature of thermal effects.
\end{itemize}

\subsection{Melon topology}\label{Th-QFTmelon}

The contribution of the melon topology to the free energy density in $d$ space dimensions in the imaginary time formalism of Th-QFT is
\be\label{melonTh-QFT}
f_{\rm m}~=~\begin{tikzpicture}[line width=1.1 pt, scale=.6, baseline=(current bounding box.center)]

		\draw (0,0) circle (1) ;
		\draw (0,0) ellipse (.4 and 1) ;
	
 \end{tikzpicture}~=~\frac{\lambda^2}{48}\SumInt_{P,Q,R} \frac 1{P^2\,Q^2\,R^2\,(P+Q+R)^2}\,,
\ee
where $\SumInt_P$ is a standard shorthand for $\beta^{-1}\sum_{m\in {\mathbb Z}}\int \frac{\dd^d p}{(2\pi)^3}$, while $P^2={\bm p}^2+(2\pi m \beta^{-1})^2$. Using the Poisson summation formula repeatedly one gets
\be\label{melone}
f_{\rm m}=\frac{\lambda^2}{48}\left(\,\prod_{i=1}^3\int\frac{\dd^dk_i\dd\omega_i}{(2\pi)^{d+1}}\sum_{\ell_i=-\infty}^{+\infty}\frac{e^{-i\ell_i\omega_i\beta}}{\omega_i^2+{\bm k}_i^2}\,\right)\frac{1}{(\omega_1+\omega_2+\omega_3)^2+({\bm k}_1+{\bm k}_2+{\bm k}_3)^2}\,.
\ee
The expression in \eq{melone} has a sum over three integers $\ell_1,\ell_2,\ell_3$, and one can split it into four types of contribution, defined by the number of $\ell$s that are set equal to zero. The term with $\ell_i=0$ for all $i$ is the contribution to the cosmological constant. Terms with 2, 1 and no $\ell$ set to zero contribute respectively to Eqs.~(\ref{1to1}), (\ref{Fm2}) and (\ref{Fm3}).

To see an example of how the matching to DMB works, we choose $\ell_1>0$, $\ell_2<0$ and $\ell_3=0$. With only one $\ell$ set to zero, we expect it to contribute to \eq{Fm2}. The key to go to a real time expression is to perform the $\dd\omega_1$ and $\dd\omega_2$ integrals with the residue theorem, closing the contour in respectively the lower and upper half plane as dictated by the signs of $\ell_{1,2}$. By taking the residue at the poles in $\omega_1=-i|{\bm k}_1|$, $\omega_2=+i|{\bm k}_2|$ and doing the sum over $\ell_1$ and $\ell_2$, it is possible, after having performed an inverse Wick rotation $\omega_3\to- i  (k_0)_3$, to rewrite
\be\label{fm><=}
f_{\rm m}\big|_{*}=\frac{\lambda^2}{48}\int \dd\Phi_{d,{\bm k}_1}^{\rm th}\dd\Phi_{d,{\bm k}_2}^{\rm th}\int\frac{\dd^{d+1}k_3}{i(2\pi)^{d+1}k_3^2 \,(k_3+k_1-k_2)^2}=\frac{\lambda^2}{48}\int \dd\Phi_{d,{\bm k}_1}^{\rm th}\dd\Phi_{d,{\bm k}_2}^{\rm th}B_d(-u)\,,
\ee
where we identified $u=(k_1-k_2)^2$ in the forward limit, and the $*$ is meant to remind that we are doing the computation with a restriction on $\ell_i$.

In order to match \eq{Fm2to2} we need a factor of 12, that should correspond to the number of distinct regions of integration in \eq{melonTh-QFT} that produce a contribution identical to \eq{fm><=}. 
To get this number we have to first count the number of ways in which the melon diagram can be cut twice, which is $\binom 42=6$. Then one needs to decide which side of each cut leg goes to the outgoing state, and which side to the incoming state. In the example we just considered, this corresponds to the choice of {sign} of $\ell_1$ and $\ell_2$. There are two choices to make, one for each cut leg. Out of the four possibilities, two produce a contribution like \eq{fm><=}, or Fig.~\ref{fig:melon}$(a)$, while the remaining two give an analogous integral with $B_d(-s)$ instead of $B_d(-u)$, corresponding to Fig.~\ref{fig:melon}$(b)$. All in all, we get the desired factor of 12.
To fill in the details, see for example \cite{Schubring:2024yfi}.

\section{Cancelation of imaginary parts}\label{appim}

Next we show  the cancellation of imaginary contributions among the diagrams with melon topology. The reality of $F$, as computed with \eq{master}, can also be seen directly from the formula, because the ${\rm Tr} \ln U$ is purely imaginary when $U$ is a unitary operator.

\begin{figure}[t] 
   \centering
   
   \begin{tikzpicture}[line width=1.3 pt, scale=.7, baseline=(current bounding box.center)]
   
	\draw (-.7,0) -- (-.3,1) -- (-.7,2) ;
	\draw (.7,0) -- (.3,1) -- (.7,2) ;
	\draw (-.7,2) -- (-.3,3) -- (-.7,4) ;
	\draw (.7,2) -- (.3,3) -- (.7,4) ;
	\draw[fill=gray!20] (0,1) ellipse (.7 and .4);
	\draw[fill=gray!20] (0,3) ellipse (.7 and .4);
	\draw[gray!40] (-1.5,0) -- (1.5,0) ;
	\draw[gray!40] (-1.5,2) -- (1.5,2) ;
	\draw[gray!40] (-1.5,4) -- (1.5,4) ;
	\node at (-1.9,0) {$a$} ;
	\node at (-1.92,2.1) {$a'$} ;
	\node at (-1.9,4) {$a$} ;
	\node at (0,1) {$c$};
	\node at (0,3) {$c$};

	\begin{scope}[xshift=8cm]
	\draw (-.75,.5) -- (-1.3,2) -- (-.75,3.5) ;
	\draw (.75,.5) -- (1.3,2) -- (.75,3.5) ;
	\draw (-2.25,.5) -- (-1.7,2) -- (-2.25,3.5) ;
	\draw (2.25,.5) -- (1.7,2) -- (2.25,3.5) ;
	\draw[fill=gray!20] (-1.5,2) ellipse (.7 and .5);
	\draw[fill=gray!20] (1.5,2) ellipse (.7 and .5);
	\draw[gray!40] (-3,.5) -- (3,.5) ;
	\draw[gray!40] (-3,3.5) -- (3,3.5) ;
	\node at (-3.4,.5) {$a$} ;
	\node at (-3.4,3.5) {$a$} ;
	\node at (-2.25,0) {\scriptsize 1} ;
	\node at (-.75,0) {\scriptsize 2} ;
	\node at (.75,0) {\scriptsize 3} ;
	\node at (2.25,0) {\scriptsize 4} ;
	\node at (-2.25,4) {\scriptsize 3} ;
	\node at (-.75,4) {\scriptsize 4} ;
	\node at (.75,4) {\scriptsize 1} ;
	\node at (2.25,4) {\scriptsize 2} ;
	\node at (-1.5,2) {$c$};
	\node at (1.5,2) {$c$};
	
	\end{scope}
	
   \end{tikzpicture}

   \caption{Diagrams that contribute at $O(\lambda^2)$ to the imaginary part of the free energy. The first comes from expanding $\ln\big(1-2\pi i\delta(E-H_0) T\big)$ in \eq{master} to second order about the free theory. The second comes from a single insertion of $T$, where the resulting amplitude is disconnected in the usual sense and becomes connected only after the identification of initial and final states. The so obtained topology of these diagrams is of the melon type.}
   \label{fig:purelyIm}
\end{figure}
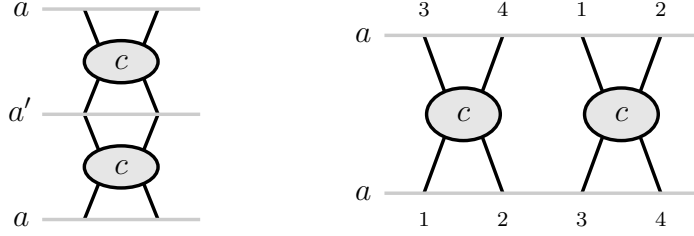

Imaginary parts come from four distinct sources. We are going to list them and evaluate them in turn.
\begin{enumerate}
\item The first contribution comes from diagram $(b)$ of Figure~\ref{fig:melon}, which corresponds to the term proportional to $B_d(-s)$ in \eq{melon22}. It can be computed directly using ${\rm Im}[\ln(-s)]=-\pi$, and is equal to
\begin{align}\label{im}
{\rm Im}\big[F_{(\ref{melon22})}\big]&=\frac{\lambda^2L^3}{64\pi}\,\mu_{\rm th}^2\,.
\end{align}
The imaginary contribution has support in the region of kinematic space where virtual particles are produced on-shell, as dictated by the Cutkosky rule. In formulas, this can be expressed as
$
{\rm Im}\la 12|T|12\ra=\frac{L^3}{2 \cdot2!}\int \dd\Phi_{1'}\dd\Phi_{2'}\,M_{12\to1'2'}\,M_{1'2'\to12}\,(2\pi)^4\delta^{(4)}(p'-p)
$,
where the first factor $\frac12$ is dictated by the optical theorem ${\rm Im}\,T=\frac12 T^\dagger T$, while the $\frac1{2!}$ is there because $1'$ and $2'$ are identical, and where $M$ is obtained from $T$ by stripping the Dirac delta of momentum conservation off. The expression for the imaginary part of a $T$-matrix element allows to write
\be\label{Im22}
{\rm Im}\big[F_{(\ref{melon22})}\big]=\frac{L^3}8\int\dd\Phi_1^{\rm th}\dd\Phi_2^{\rm th}\,\int \dd\Phi_{1'}\dd\Phi_{2'}\,M_{12\to1'2'}\,M_{1'2'\to12}\,(2\pi)^4\delta^{(4)}(p'-p)\,.
\ee
Notice that, among the four copies of one-particle phase space measures, two are thermal and two are not.
\item The second contribution comes from diagram $(c)$ of Figure \ref{fig:melon} and its two companions (those that are obtained by permuting 123). The imaginary part comes from the kinematic region where $(p_1+p_2-p_3)^2= 0$. On the kinematic poles the amplitude factorises into a product of $2\to2$ tree-level amplitudes, so we can write
\be\label{Im33}
{\rm Im}\big[F_{(\ref{T123tree})}\big]=3\times\frac{L^3}{12}\int\dd\Phi_1^{\rm th}\dd\Phi_2^{\rm th}\dd\Phi_3^{\rm th}\int \dd\Phi_{1'}\,M_{12\to1'3}\,M_{1'3\to12}\,(2\pi)^4\delta^{(4)}(p'-p)\,,
\ee
where the factor of 3 comes about because we added the contribution from the regions where either $(-p_1+p_2+p_3)^2= 0$ or $(p_1-p_2+p_3)^2= 0$. The structure of \eq{Im22} and \eq{Im33} is identical, except for the presence of three, instead of two, Bose-Einstein density factors in the latter (substituting $3$ with $2'$ in (\ref{Im33}) makes it apparent).
\item Until here we have considered only contributions to the free energy that are proportional to one power of the $T$-matrix, despite the fact that \eq{master} contains terms with an arbitrary number of $T$ insertions. At $O(\lambda^2)$ we surely need to include, in \eq{master}, contributions of the form $\int \la 12|T|1'2'\ra\la1'2'|T|12\ra$, where each matrix element is $O(\lambda)$ (see the leftmost diagram of Figure~\ref{fig:purelyIm}). On top of this, we should also include all the `histories' that differ from it by an arbitrary number of free particle propagations, with the condition that the resulting history has a trace connected diagram. The result of the resummation was given in \cite{Baratella:2024sax}, and gives a contribution which is purely imaginary at $O(\lambda^2)$
\be
F_{2\to2\to2}=
-i\,\frac{L^3}8\int\dd\Phi_1^{\rm th}\,\,\dd\Phi_2^{\rm th} \,\,\dd\Phi_{1'}^{\rm th}\,\,\dd\Phi_{2'}^{\rm th}\,\,e^{\beta(|\bm k_1|+|\bm k_2|)}M_{12\to1'2'}\,M_{1'2'\to12}\,(2\pi)^4\delta^{(4)}(p'-p)\,.
\ee
\item Finally, there is another type of contribution which arises from one insertion of $T$, and whose leading contribution in $n$ is a $4\to 4$ process. It is disconnected in the usual S-matrix sense, as it decomposes into two clusters.
However, when initial and final states are properly identified the amplitude becomes trace-connected and has the melon topology (see the rightmost diagram of Figure~\ref{fig:purelyIm}). Its contribution to the free energy can be expressed as
\be\label{4to4}
F_{4\to4}=
3\times i\,\frac{L^3}{24}\int\dd\Phi_1^{\rm th}\,\dd\Phi_2^{\rm th} \,\dd\Phi_{3}^{\rm th}\,\dd\Phi_{4}^{\rm th}\,M_{12\to34}\,M_{34\to12}\,(2\pi)^4\delta^{(4)}(p'-p)\,,
\ee
where the factor 3 comes from the number of distinct clusterings of $1234\to1234$ that result in a melon-type topology.
\end{enumerate}
The four contributions cancel among themselves non-trivially. To see this at the integrand level, it is enough to relabel $3\to 2'$ in \eq{Im33}, $3,4\to 1',2'$ in \eq{4to4}, symmetrise over $1'$ and $2'$ in \eq{Im33} and finally invoke the identity\footnote{I thank Emanuele Gendy for figuring this out at a stage in which the consistency of \eq{master} beyond leading order was not obvious to us.}
\be
n_B(|\bm k_1|)^{-1}n_B(|\bm k_2|)^{-1}+\big[n_B(|\bm k_1|)^{-1}+n_B(|\bm k_2|)^{-1}\big]+1-e^{\beta |\bm k_1|}e^{\beta |\bm k_2|}=0\,.
\ee 
We conclude that, to $O(\lambda^2)$, physical information about the free energy content of the theory only comes from the one-loop $2\to 2$ amplitude and from the $3\to 3$ connected tree-level amplitude, while diagrams of Figure~\ref{fig:purelyIm} are only there to cancel imaginary parts.

\section{Free energy at NNLO: long-caterpillar topology}\label{app:grid}

We report here the value of all seed diagrams times the appropriate Bose-Einstein density factors. Out of the six possible orderings, we consider here only $F_{t_{\rm rb}<t_{\rm r}<t_{\rm b}}$ and $F_{t_{\rm r}<t_{\rm rb}<t_{\rm b}}$. The total is obtained as
\be
F=\bigg(2\,F_{t_{\rm rb}<t_{\rm r}<t_{\rm b}}+F_{t_{\rm r}<t_{\rm rb}<t_{\rm b}}\bigg)+ \,{\rm r}\leftrightarrow {\rm b}\,.
\ee

\subsection{$t_{\rm rb}<t_{\rm r}<t_{\rm b}$}
With two singular denominators:
\be
\frac{(AA)}{n_{\rm r}n_{\rm b}}=\frac{(AB)}{n_{\rm r}n_{\rm b}^2}=\frac{(BA)}{n_{\rm r}^2n_{\rm b}}=\frac{(BB)}{n_{\rm r}^2n_{\rm b}^2}=\frac{\beta^2}6\,.
\ee
With one singular denominator:
\be
\frac{(AC)}{n_{\rm r}n_{\rm b}}=\frac{(AD)}{n_{\rm r}n_{\rm b}^2}=\frac{(BC)}{n_{\rm r}^2n_{\rm b}}=\frac{(BD)}{n_{\rm r}^2n_{\rm b}^2}=\frac\beta{4|\bm q_{\rm b}|}-\frac1{8|\bm q_{\rm b}|^2}\,,
\ee
\be
\frac{(AE)}{n_{\rm r}n_{\rm b}^2}=\frac{(BE)}{n_{\rm r}^2n_{\rm b}^2}=-\frac\beta{4|\bm q_{\rm b}|}-\frac1{8|\bm q_{\rm b}|^2}\,.
\ee
With no singular denominator:
\be
\frac{(CA)}{n_{\rm r}n_{\rm b}}=\frac{(CB)}{n_{\rm r}n_{\rm b}^2}=\frac{(DA)}{n_{\rm r}^2n_{\rm b}}=\frac{(DB)}{n_{\rm r}^2n_{\rm b}^2}=\frac{(EA)}{n_{\rm r}^2n_{\rm b}}=\frac{(EB)}{n_{\rm r}^2n_{\rm b}^2}=\frac1{4|\bm q_{\rm r}|^2}\,,
\ee
\be
\frac{(CC)}{n_{\rm r}n_{\rm b}}=\frac{(CD)}{n_{\rm r}n_{\rm b}^2}=\frac{(DC)}{n_{\rm r}^2n_{\rm b}}=\frac{(DD)}{n_{\rm r}^2n_{\rm b}^2}=\frac{(EE)}{n_{\rm r}^2n_{\rm b}^2}=\frac1{4|\bm q_{\rm r}|(|\bm q_{\rm r}|+|\bm q_{\rm b}|)}\,,
\ee
\be
\frac{(CE)}{n_{\rm r}n_{\rm b}^2}=\frac{(EC)}{n_{\rm r}^2n_{\rm b}}=\frac{(ED)}{n_{\rm r}^2n_{\rm b}^2}=\frac{(DE)}{n_{\rm r}^2n_{\rm b}^2}=\frac1{4|\bm q_{\rm r}|(|\bm q_{\rm r}|-|\bm q_{\rm b}|)}\,.
\ee

\subsection{$t_{\rm r}<t_{\rm rb}<t_{\rm b}$}
We limit here to the upper triangular matrix
\be
\begin{pmatrix}AA & AB & AC & AD & AE \\  & BB & BC& BD& BE \\ && CC & CD & CE \\ &&& DD & DE \\ &&&& EE\end{pmatrix}
\ee
since the remaining objects can be simply obtained by symmetrisation in r and b.

With two singular denominators:
\be
\frac{(AA)}{n_{\rm r}n_{\rm b}}=\frac{(AB)}{n_{\rm r}n_{\rm b}^2}=\frac{(BB)}{n_{\rm r}^2n_{\rm b}^2}=\frac{\beta^2}6\,.
\ee
With one singular denominator:
\be
\frac{(AC)}{n_{\rm r}n_{\rm b}}=\frac{(AD)}{n_{\rm r}n_{\rm b}^2}=\frac{(BC)}{n_{\rm r}^2n_{\rm b}}=\frac{(BD)}{n_{\rm r}^2n_{\rm b}^2}=\frac\beta{4|\bm q_{\rm b}|}-\frac1{8|\bm q_{\rm b}|^2}\,,
\ee
\be
\frac{(AE)}{n_{\rm r}n_{\rm b}^2}=\frac{(BE)}{n_{\rm r}^2n_{\rm b}^2}=-\frac\beta{4|\bm q_{\rm b}|}-\frac1{8|\bm q_{\rm b}|^2}\,.
\ee
With no singular propagator (notice the global signs):
\be
\frac{(CC)}{n_{\rm r}n_{\rm b}}=\frac{(CD)}{n_{\rm r}n_{\rm b}^2}=\frac{-(CE)}{n_{\rm r}n_{\rm b}^2}=\frac{(DD)}{n_{\rm r}^2n_{\rm b}^2}=\frac{-(DE)}{n_{\rm r}^2n_{\rm b}^2}=\frac{(EE)}{n_{\rm r}^2n_{\rm b}^2}=\frac1{4|\bm q_{\rm r}||\bm q_{\rm b}|}\,.
\ee

\section{Cancelation of spurious $\varepsilon$-finite terms at NNLO}
\label{ic}

We complete the computation of $O(\lambda^3)$ corrections to the free energy by proving that terms that were omitted in
\eq{toctwo} and \eq{tocone} cancel. 

From \reef{toctwo}, one might naively expect that 
\be
\Delta(E)\equiv(G_a-\bar G_a)\left(-\frac12 G_a+\frac 13 (G_a-\bar G_a)\right)(G_b-\bar G_b)
\ee
vanishes because it is   proportional to $\delta(E-E_a)\delta(E-E_b)$, with $E_a\neq E_b$. However, this is not true, because the identification $(G_a-\bar G_a)=-2\pi i\delta(E-E_a)$ is only valid in the limit $\varepsilon\to 0$, a limit which  must be taken only at the very end of the calculation. Following the correct procedure one then finds
\be\label{Deltaint}
-\frac{1}{2\pi i}\,\lim_{\varepsilon\to 0}\int \dd E \,\Delta(E) \,f(E)=-\frac16 f(E_a)\frac1{(E_a-E_b)^2}
\ee
where $f(E)$ is a regular function. To understand where this piece comes from, notice first that $(G_a-\bar G_a)^2$ is a function that has a peak at $E=E_a$ with width $\varepsilon$ and height $\frac1{\varepsilon^2}$. Second, note that $G_b-\bar G_b$, which is peaked at $E=E_b$ with width $\varepsilon$ and height $\frac1{\varepsilon}$, has tails that go like $\frac\varepsilon{(E-E_a)^2}$.
The region where the tail of $G_b-\bar G_b$ overlaps with the bump of $(G_a-\bar G_a)^2$ one gets a peak of width $\varepsilon$ and height $\frac1{\varepsilon^2}\varepsilon$, which gives a contribution to the integral in (\ref{Deltaint}) that remains finite for $\varepsilon\to 0$. If it was not for the long tail of $G_b-\bar G_b$, the integral would give zero like naively expected.

Similar contributions appear also in \eq{tocone}. Taking the left-hand-side of the equation and integrating it in $\dd E$ times a regular function, one has
\be\label{corr2}
\frac{-1}{2\pi i}\,\lim_{\varepsilon\to 0}\int \dd E \,\,{\rm LHS}(\ref{tocone}) \,f(E)=\frac{f(E_a)}{(E_a-E_b)^2}+\frac13 \frac{f(E_b)}{(E_a-E_b)^2}\,.
\ee
The first contribution corresponds to the one on the RHS of \eq{tocone}, that we already accounted for, while the second is new. 
We emphasise that in the last term of \reef{corr2}, $f(E_b)$ and not $f(E_a)$ appears; while $f(E_a)$ appears in \reef{Deltaint} and the first term of \reef{corr2}.

Now one has to take the seed diagrams of Figure \ref{fig:paluV3} and account for the corrections of \eq{Deltaint} (for diagrams in the second row) and \eq{corr2} (for diagrams in the third row), and multiply by the appropriate power of $n_B$. All in all one has
\begin{align}\label{corr}
&f_{\rm correction}=\,\,\lambda^3 \mu_{\rm th}^3\int \dd \Phi_{\bm k} \,\frac1{16|\bm k|^4}\Bigg\{ -\frac16\,\big[ 2\,n_B +6\,n_B^2+4\,n_B^3\big] \\
&+\frac13\,\big[ 3\,e^{-2\beta|{\bm k}|}n_B +3\,e^{-2\beta|{\bm k}|}n_B^2+e^{-2\beta|{\bm k}|}n_B^3+e^{2\beta|{\bm k}|}n_B^3\big]\Bigg\}=\lambda^3 \mu_{\rm th}^3\int \dd \Phi_{\bm k} \,\frac1{16|\bm k|^4}\,\frac{-1}3\,e^{-2\beta|{\bm k}|}\,,\nonumber
\end{align}
where the extra powers of $e^{-\beta|{\bm k}|}$ in the second line come from terms proportional to $e^{-\beta E_b}$ instead of $e^{-\beta E_a}$. The end result is not zero, but it is proportional to simply $e^{-2\beta|{\bm k}|}$ (all the complicated dependence on $e^{-\beta|{\bm k}|}$ through $n_B$ simplified away). Notice however that we are missing a diagram that receives a contribution from the correction terms of this Appendix, and is equal to
\be\label{vacuum}
\begin{tikzpicture}[line width=1.1 pt, scale=.6, baseline=(current bounding box.center)]

		\draw (0,0) to [bend left=60] (0,2);
  		\draw (0,0) to [bend right=60] (0,2);
		\filldraw (0,0) circle (4pt);
		\filldraw (0,2) circle (4pt);
		\filldraw (.5,1) circle (4pt);
		\node at (2,1) {$\bigg|_{\rm correction}$};
	
 \end{tikzpicture}~=~\lambda^3 \mu_{\rm th}^3\int \dd \Phi_{\bm k} \,\frac1{16|\bm k|^4}\,\frac{1}3\,e^{-2|{\bm k}|}\,,
\ee
which exactly cancels with \ref{corr}. This is the first example of a contribution that is thermal in the ${\bm k}$ momentum despite superficially the loop of ${\bm k}$ looks like a vacuum loop  -- note that it is not a  vacuum diagram, but looks like so in this notation, see Fig.~\ref{fig:palu} and explanations of it. The contributions from the diagram in \eq{vacuum} that have no power of $e^{-\beta|{\bm k}|}$ are expected to cancel along the lines of the introduction to Section~\ref{sec:singular}.

\bibliographystyle{unsrt}
\bibliography{references}

\end{document}